\documentclass{article}[12pt]
 \usepackage{a4wide}
\usepackage[pdftex]{graphicx}
\usepackage{subfigure}
\usepackage{amsmath}
\usepackage{epsfig}
\usepackage{amssymb}
\usepackage{latexsym}
\usepackage[cmtip,arrow]{xy}
\usepackage{subfigure}
\usepackage{epsfig}
\usepackage{amssymb}
\usepackage{latexsym}
\newtheorem{theorem}{Theorem}
\newtheorem{corollary}{Corollary}
\newtheorem{definition}{Definition}
\newtheorem{lemma}{Lemma}

\newtheorem{proposition}{Proposition}
\usepackage{pb-diagram,pb-xy}
\newtheorem{remark}{Remark}
\newenvironment{proof}{{\bf Proof.}}{\hfill $\square$}
\newcommand{\reach}{\textrm{reach}}

\newcommand{\np}{\mathit{np}}
\newcommand{\NP}{\mathit{Np}}

\newcommand{\C}{\mathcal{C}}
\newcommand{\R}{\mathbb R}
\newcommand{\RC}{\mathcal{R}_{\C}}

 \newcommand{\MC}{\mathcal{M}_{\C}}
 \newcommand{\WMC}{\widetilde{\mathcal{M}}_{\C}}

\newcommand{\OO}{\mathcal{O}}

\newcommand{\SSS}{\mathcal{S}}
\newcommand{\OC}{\OO_{\C}}

\newcommand{\RR}{\mathcal{R}}
\newcommand{\SC}{\SSS_{\C}}
\newcommand{\vor}{{\rm Vor}}
\newcommand{\lift}{\mathrm{lift}}

\newcommand{\LSC}{\lift(\SC)}
\newcommand{\MA}{\text{MA}} 
\newcommand{\VD}{\mathrm{VorSkel}}

\newcommand{\LLL}{\mathcal{L}}

\title{Geometric Tomography\\ With Topological Guarantees
}

\author{Omid Amini$^*$ \hspace{2mm} Jean-Daniel Boissonnat$^{\dagger}$ \hspace{2mm} Pooran Memari$^{\ddagger}$}
\date {}
\begin{document}
\maketitle

\section*{abstract}{\footnote {
An extended abstract of this work appeared in the proceedings of the 26th Annual Symposium on Computational Geometry (SoCG), 2010. \\
$^*$ CNRS-DMA, \'Ecole Normale Sup\'erieure, France, \\
$^{\dagger}$ INRIA Sophia Antipolis - M\'editerran\'ee, France,\\
$^{\ddagger}$ CNRS-T\'el\'ecom ParisTech, France.  Correspondance: memari@telecom-paristech.fr\\
This work was done while this author was Ph.D student at INRIA Sophia Antipolis - M\'editerran\'ee, France.}}
We consider the problem of reconstructing a compact
3-manifold (with boundary) embedded in $\mathbb{R}^3$ from its cross-sections $\SSS$ with a given
set of cutting planes $\mathcal P$ having arbitrary orientations. In this paper, we analyse a very natural reconstruction strategy: a point $x \in \R^3$ belongs to the reconstructed object if (at least one of) its nearest point(s) in $\mathcal P$ belongs to $\SSS$. We prove that under appropriate sampling conditions, the output of such an algorithm preserves the homotopy type of the original object. Using the homotopy equivalence, we also show that the reconstructed object is homeomorphic (and isotopic) to the original object. This is the first time that 3-dimensional shape reconstruction from cross-sections comes with theoretical guarantees.

%
%


\section{Introduction}

\begin{figure}[htb] 
 \begin{center}
\includegraphics[width=0.9\linewidth]{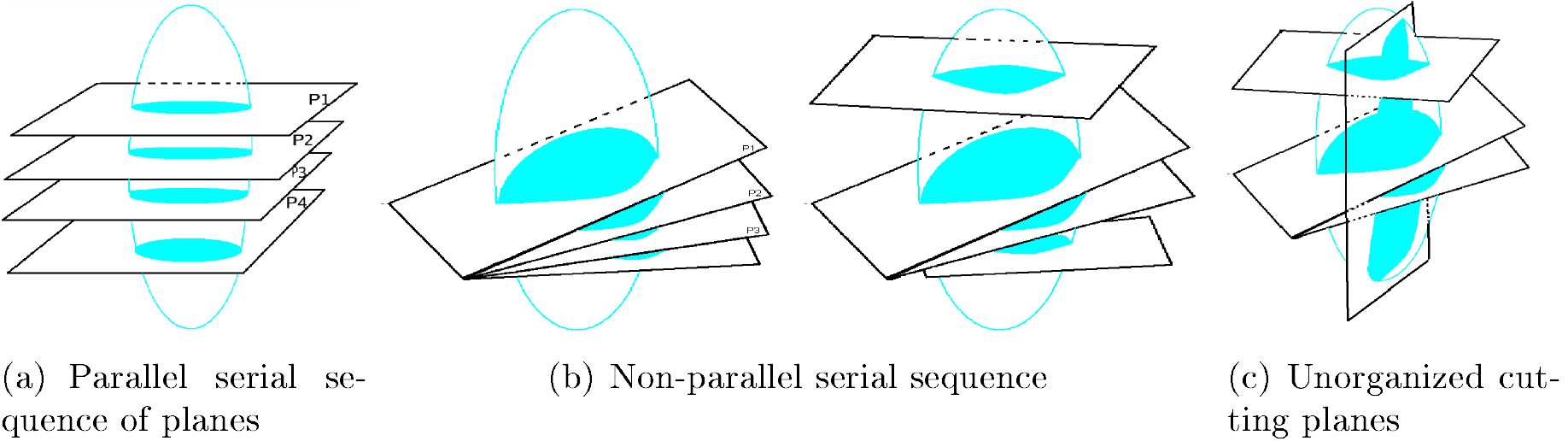}
 \end{center}\vspace{-2mm}
 \caption{Different cross-sections configurations.}
 \label{planes-positions}
 \end{figure}

Geometric tomography consists of reconstructing a 3-dimensional object from 2-dimensional information such as projections or sections. 
In this paper, we are interested in the reconstruction of a 3D object from its intersections with a set of planes, called cross-sections.
Different configurations can be considered for the cutting planes, as it is illustrated in Figure~\ref{planes-positions}. A so-called {\em serial sequence} of planes partitions the space into several {\em slices}, each bounded by a pair of cutting planes. Many solutions to the reconstruction problem from parallel or serial cross-sections have been proposed, see \cite{PooranThesis} for a survey. However, this paper focuses on the harder and more general case of arbitrarily oriented (multi-axial) planes, which has received much less
attention.
 
 \medskip
\paragraph{\bf Motivations} 
The need for reconstructions from unorganized cross-sections is a result of the advances in medical imaging
technology, especially in ultrasound tomography. In this context, the purpose is to construct a 3D model of an organ from a collection of ultrasonic images. When the images are provided by free-hand ultrasound devices, the cross-sections of the organ belong to planes that are not necessarily parallel. 

Another recent potential domain of application is the SPIM (Single Plane Illumination Microscope) which is an original microscope developed by the European Molecular Biology Laboratory in Germany around 2004 \cite{HS+04}. It allows to study functional aspects of living specimens and raises several challenges in different fields such as optics, image processing, mathematical modeling and scientific computing. Contrarily to more mature imaging technologies like confocal or multi-photon microscopes, SPIM does not rely on scanning a single focal point throughout the sample: a light sheet is produced and is used to illuminate a slice of the whole sample. This light sheet is positioned in the focal plane of a microscope objective, and the (cutting) plane can be translated and rotated. Since the slices provided by SPIM have different orientations and positions, the reconstruction of 3D structures using these data is a difficult inverse problem that is still open.

The hope is that the topological analysis provided in this paper will be an initial theoretical step toward robust and efficient reconstruction algorithms for these applications.   

Another motivation for the reconstruction from cross-sections is from a proper geometrical point of view. The $d$-dimensional framework of the problem consists of reconstructing a $d$-dimensional shape from its intersections with hyperplanes, of dimension $(d-1)$. Finding efficient methods for this reconstruction problem may be used in 
{\em topological coding of objects} or in {\em dimension reduction strategies}. Adding {\em time} as the fourth dimension to three-dimensional space, the fourth dimensional variant of the reconstruction problem can be applied to {\em moving objects} (for example the beating heart). In the sense that the movement of the object can be reconstructed knowing its position at some discrete values of time. Understanding the geometry of the 3D problem is a preliminary step toward more theoretical results for higher dimensions.

\medskip
\vspace{-3mm}
\paragraph{\bf Previous work on reconstruction from parallel cross-sections}
The classical case of parallel cross-sections has been considered widely in the past. The existing techniques for reconstruction from parallel sections may be roughly classified
into two main groups: 

\medskip

\noindent {\bf a. Surface extraction methods:} We could cite \cite{CL96}, \cite{FD02}, \cite{B05} and \cite{Mal06} for implicit surfaces, as well as \cite{PK96}, \cite{PT96} and \cite{P05} which use B-spline surface approximations. In particular, Marching Cubes (introduced in~\cite{LC87}, and (corrected and) completed in~\cite{NH91}) is widely used namely in medical imaging softwares, but no theoretical guarantees are available to make the resulting reconstruction appropriate for diagnosis.
\medskip

\noindent {\bf b. Interpolation approaches:} Since Keppel's work~\cite{K75} on interpolation between parallel polygonal contours, numerous algorithms were introduced for parallel inter-slice interpolation.
This was soon followed by Fuchs et al. in \cite{FK77}, with a minimal surface area solution. Other criteria have included minimal length of the next edge in \cite{CS78}, optimal verticality of the edges in \cite{GD82}, minimal radii of circumscribed circles of triangles  in Meyers' Ph.D thesis \cite{M94}, and minimal sum of absolute value of angle between the contour edge parts of successive triangles \cite{WW94}. Most of these interpolation techniques failed when confronted with contours
which cannot be interpolated without the addition of extra vertices (branching situation). Among the earliest methods able to handle branching between sections we can cite \cite{CS78}, \cite{S81}, \cite{B88}, and \cite{EPO91}, which were then followed by different other approaches, namely \cite{BG93}, \cite{CP94}, \cite{BCL96}, \cite{BS96}, \cite{BGLS04}, which attempted to handle the most general situation in which the geometries and topologies of the contours in every slice are unrestricted (still in the case of parallel cross-sections).  As examples of type of different methods we could cite methods based on parametric domain triangulation such as \cite{O+96}, \cite{CP01}, \cite{W+06}, or Delaunay-based methods such as \cite{BG93}, \cite{TCC94}, \cite{DP97} and \cite{CD99}.


\medskip

Since non of the above mentioned work treat the question of topological guarantees for the results, which is the primary purpose of our paper, we remain at this level of presentation and simply refer to the PhD thesis of the third author~\cite{PooranThesis} for a more complete survey.


\medskip

\paragraph{\bf Previous work on reconstruction from unorganized (non-parallel) cross-sections} Despite all these potential applications, it is only very recently that reconstruction
from unorganized cross-sections has been considered. A very first work by Payne and Toga~\cite{PT94}  was restricted to easy cases of reconstruction that do not require branching between sections. In \cite{BG93}, Boissonnat and Geiger proposed a Delaunay-based algorithm for the case of serial planes, that has been generalized to arbitrarily oriented planes in \cite{DP97} and \cite{BM07}. 
In \cite{turk99}, Turk and O'Brien presented a reconstruction method using variational implicit functions which combines the two steps
of building implicit functions and interpolating between them. Some more recent work~\cite{JW+05,LB+08} can handle the case of multi label sections (multiple materials). Barequet and Vaxman's work \cite{BV09} extends the work of \cite{LB+08} and can handle the case where the sections are only partially known. \cite{BVG11} can be cited as the most recent work that appeared in this area. In addition to the cases handled by previous work, their method deals with the {\em online} (interactive) variant of the problem, typical in freehand ultrasound scanning applications, in
which the reconstruction is updated using the slices that are given gradually over time. 

None of these methods provides a topological study of the reconstruction. The only existing
results studying the topology of the reconstructed object are restricted to
the 2D variant of the problem (\cite{SBG06}, \cite{MB08}). We will also study the 2D problem in Section~\ref{sec:2d} of this paper, and compare our method here to our previous one in~\cite{MB08}.

\medskip

\paragraph*{\bf Methodology analyzed in this paper}  
Most of the previous work for the case of parallel cross-sections are composed of two steps: correspondence and interpolation. The correspondence step consists in finding the right connectivity between different sections. It is very common to use a projection based strategy to perform the connections, and naturally the direction orthogonal to the planes is the most common. Most of such methods use (directly or indirectly) the simple idea of connecting two sections if their orthogonal projections overlap. Moreover, even different methods such as implicit functions seem to perform almost the same connectivity rule, since in most of the cases, the function defined between the slices promote the orthogonal direction around each cross-section.     

In this paper, we analyze a natural generalization of this idea for the case of nonparallel sections, proposed by Liu et al. in \cite{LB+08}. We prove that under appropriate sampling
conditions, the performed connection between the sections is coherent with the connectivity structure of the object and the proposed reconstructed object is homeomorphic to the
object. 

\medskip

\paragraph*{\bf Contributions}
We make the first geometric analysis of the reliability and validity of reconstruction
methods from cross-sectional data, by analyzing the most
natural way to connect the sections, which is a generalization of the classical overlapping
criterion between the orthogonal projections of parallel sections. We provide sampling conditions for the set of cutting planes, under which the reconstructed object is
isotopic to the original unknown object. As we will see, knowing only a lower bound for the minimum local feature size (defined in the next section) of the (unknown) object is sufficient to verify these sampling conditions.  

No reconstruction method from cross-sections came with theoretical guarantees before this work. Even in the case of parallel cross-sections, no formal analysis and guarantees have been obtained. This contrasts with the problem of reconstructing a shape from unorganized
point sets, which is now well understood both theoretically and
practically (via e.g. closed ball property and epsilon sampling). This work is the first, but a crucial, step to close the gap between the reconstruction problem from cross-sections and the classical problem of reconstruction from sample points. 

\medskip
 
\paragraph*{\bf Sectional Data Compared to Point Cloud Data.}

We note that the reconstruction problem from cross-sections could be
also viewed as an instance of point cloud reconstruction problem if we
consider a sample of the given cross-sections or their contours.
However, the following basic example shows that the sampling conditions
we propose in this paper are indeed much better adapted to the
reconstruction problem in this context, when compared to classical
sampling conditions for point clouds.
 
Consider the Voronoi diagram of the sample points as a covering of the original shape $\OO$, composed of the so-called restricted Voronoi cells. The restricted Delaunay complex is then the geometric realization of the nerve of the restricted Voronoi cells.  The nerve theorem of Leray \cite{leray} implies that if all restricted Voronoi cells are contractible then $\OO$ and the restricted Delaunay complex are homotopy equivalent. 
\begin{figure}[!htb]  
  \begin{center}
  \subfigure{\includegraphics[width=0.37\linewidth]{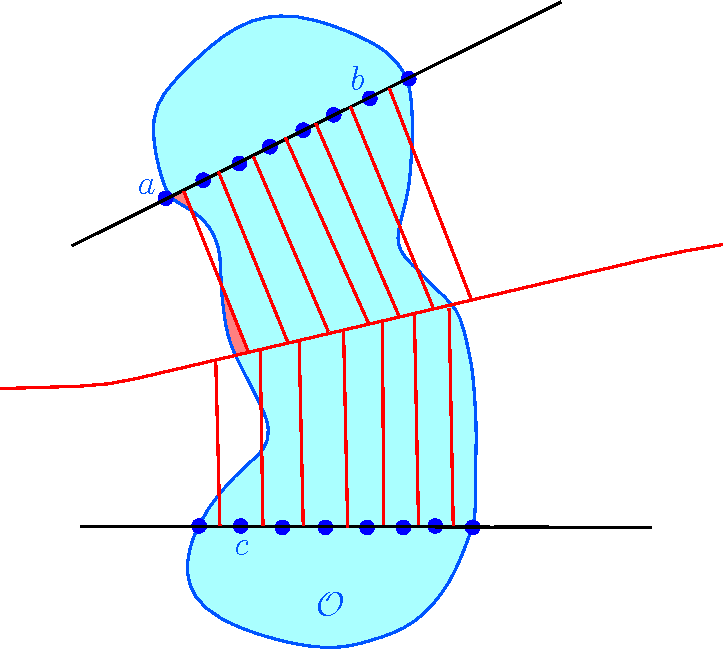}}
  \end{center}
\end{figure}

As the 2D example of the above figure shows, the Voronoi cell of the sample points on a plane $P$ form thin quadrilaterals orthogonal to $P$. To ensure the nerve theorem condition, the intersection of such a thin quadrilateral with $\OO$ should be contractible. Clearly, near to the contours (such as for points $a,b$ and $c$ in the figure), even the
connectivity of this intersection is a very strong condition and is
unverifiable in practice. On the contrary, since our sampling conditions are based on the
orthogonal projection of the section-contours, they are easily verifiable.

One justification of this is that our approach makes full use of the
additional information given by the orientation of the cutting planes,
while in point clouds reconstruction, this information is entirely lost.

\medskip
\paragraph*{\bf Overview of the results}
Consider the {\em arrangement of the cutting planes}, i.e., the subdivision
of $\R ^3$ into convex polyhedral cells induced by the cutting planes. 
Without loss of generality, we reduce the reconstruction of the original shape $\OO$ to the reconstruction of $\OC:=\OO \cap \C$
for all cells $\C$ of the arrangement. Given a cell $\C$ of the arrangement, let us define $\reach_{\C}(\OO)$ as the distance of the boundary of $\OO$ from its medial axis, in a neighborhood of $\C$ (formally defined later). We also define the height of the cell $\C$, denoted by $h_{\C}$, as the maximum distance of a point $x \in \C$ to the boundary of $\C$. Note that by definition, $h_{\C} \leq \frac{1}{2} \mathrm{diam}({\C})$.

Using the fact that a point on the boundary of $\C$ is in $\OO$ if it lies in $\SC$, we say that a point $x \in \C$ is in the reconstructed object $\RR$ if one of its {\it nearest points} in the boundary of $\C$ is in $\SC$. It is  easy to see that this generalizes the mentioned overlapping criterion for correspondence between sections. We will establish topological guarantees for the reconstruction by proposing two sampling conditions for the set of cutting planes. For the sake of simplifying the presentation in this introduction, we provide sufficient conditions rather than the formal definitions.

The first condition is verified if the set of cutting planes is sufficiently dense so that we have: 
 for any cell $\C$ of the arrangement, $h_{\C} < \reach_{\C}(\OO)$.
We show that under this condition, the connectivity between the sections in the reconstructed object $\RR$ coincides with the connectivity in the original unknown shape $\OO$. 
As we will see, this implies the homotopy equivalence between $\RR$ and $\OO$ for some particular cases: reconstruction of union of convex bodies in $\R^d$ from $d-1$ dimensional sections, and reconstruction of any compact 2-manifold in $\R^2$ from line-sections.  

However, this condition does not ensure in general the homotopy equivalence between $\RR$ and $\OO$. We need to impose a stronger condition which is verified if the cutting planes are {\em sufficiently transversal} to  the boundary of the original shape: 
 $\forall \C, h_{\C} < \frac{1}{2} \; \bigl(1 - \sin(\alpha)\bigr) \; \reach_{\C}(\OO),$ where $\alpha$ is an upper bound on the angle between the cutting planes and the normals to the boundary of $\OO$ along the sections..
\paragraph*{Main theorem} Under the two above conditions, the reconstructed object $\RR$ is homeomorphic (and isotopic) to the unknown original shape $\OO$.

\par Note that these two conditions measure the density and the transversality of the cutting planes with respect to the original shape. They are verifiable by only knowing a lower bound for the local feature size of the unknown object, which is the minimum a priori information required to estimate the order of size of the original shape to reconstruct. In practice, this a priori information about the order of local feature size is usually known, since quite often the category of shape is known in advance. For instance, for the case of medical images, one could have an estimation of the lfs from the anatomical atlas. Regarding the transversality estimation in practice, according to \cite{R03} Section 1.2.1, the quality of 2D ultrasonic images depends highly on the transversality of the cuts. 

\medskip

\paragraph*{\bf Organization of the paper} 
After this brief introduction, we provide a detailed description of the reconstructed object $\RR$. The rest of the paper will be then devoted to prove that in the general case, under two appropriate sampling conditions, the reconstructed object $\RR$ and the original object $\OO$ are homotopy equivalent, and are even homeomorphic. 

As we will see, the first sampling condition, called the {\em Separation Condition}, discussed in Section~\ref{sec:sampling}, ensures good connectivity between the sections, but does not necessarily imply the homotopy equivalence.  In order to ensure the homotopy equivalence between $\RR$ and $\OO$, a second so-called {\em Intersection Condition} is required, see Section~\ref{sec:intersection-condition}. To make the connection between the upcoming sections more clear, in Section~\ref{sec:proofoutline} we shortly outline the general strategy employed in proving the homotopy equivalence between $\RR$ and $\OO$. In Section~\ref{sec:ensure}, we provide a set of properties on the sampling of cutting planes to ensure the Separation and the Intersection Conditions. Finally, in Section~\ref{sec:homeomorphism} we show that the two shapes $\OO$ and $\RR$ are indeed homeomorphic (and even isotopic). Some preliminary notions of homotopy theory we use here are recalled in Appendix~\ref{sec:Homotopy-Preliminaries}. Let us also mention that for the sake of simplicity and better visualization, some of the figures in the paper are given in 2D. 

\section{Reconstruction Description}
\label{sec:reconstruction}

\medskip

\paragraph*{\bf Reconstruction problem} Let $\OO \subset \R^3$ be a compact 3-manifold with boundary (denoted by $\partial \OO$) of class $C^2$. 
The manifold $\OO$ is cut by a set $\mathcal P$ of the so-called cutting planes that are supposed to be in general position in the
sense that none of these cutting planes are tangent to $\partial \OO$.  For any cutting plane $P \in \mathcal P$, we are given the intersection
$\OO \cap P$. There is no assumption on the geometry or the topology of these intersections. 
The goal is to reconstruct $\OO$ from the given family $\SSS$  consisting of all the intersections of $\OO$ with the cutting planes in $\mathcal P$.

\medskip

\paragraph*{\bf Methodology} 
We know that a point $x \in \mathcal P$ belongs to the original object if and only if it belongs to $\SSS$.
The goal is now to determine whether a point $x \in \R^3$ belongs to $\OO$ or not. 
We follow a very natural reconstruction strategy: 

{\begin{center} We say that a point $x \in \R^3$ belongs to the reconstructed object if (at least one of) its {nearest point(s)} in $\mathcal P$ belongs to $\SSS$. \end{center}} 


\noindent Different distance functions (from $\mathcal P$) may be used in order to satisfy properties of interest for different applications (for example, to promote the connection between sections in the case of sparse data, or to impose a favorite direction to connect the sections, etc).
A natural idea is to use the Euclidean distance. In this case, the reconstructed object coincides with the result of the method introduced by Liu et al. in \cite{LB+08}. We will analyze this method and present appropriate sampling conditions  providing topological guarantees for the resulting reconstructed object.

\subsection{Characterization of the reconstructed object}
As said before, the definition of the reconstructed object $\RR$ is based on the distance from the set of cutting planes $\mathcal P$. Since we consider the Euclidean distance, this involves the {\em arrangement of cutting planes}, i.e., the subdivision of $\R ^3$ into convex polyhedral cells induced by the cutting planes. 
If a point $x \in \R^3$ belongs to a cell $\C$ of this arrangement, then its nearest point(s) in $\mathcal P$ belong(s) to the boundary of $\C$. Thus, we can decompose the problem into several subproblems, and reduce the reconstruction of
$\OO$ to the reconstruction of $\OC := \OO \cap \C$
for all cells $\C$ of the arrangement. 

\medskip

\paragraph*{\bf Reconstructed object in a cell of the arrangement} 
 We now focus on a cell $\C$ of the arrangement and describe the reconstructed object $\RC$ in $\C$. On each face $f$ of $\C$, the intersection
of the object $\OO$ with $f$ is given and consists of a set of connected regions called {\em sections}. By definition, the sections of a face of $\C$ are disjoint. However, two sections (on two neighbor faces of $\C$) may intersect along the intersection between their two corresponding faces. The boundary of a section $A$ is denoted by $\partial A$ and is a set of closed curves, called {\em section-contours}, that may be nested if the topology of $A$ is non-trivial. Let us write $\partial \C$ for the boundary of $\C$. In the sequel, $\SC$ denotes the union of sections of all
the faces of $\C$. 
By the definition of the reconstructed object $\RR$, we have:
 {\begin{center} A point $x \in \C$ is in the reconstructed object $\RC$ if one of its nearest points in $\partial \C$ is in $\SC$.\end{center}} 

\noindent This definition 
naturally involves the Voronoi diagram of the faces of $\C$ defined as follows.

\medskip

\begin{figure}[!htb]
   \begin{center}
 \subfigure{\includegraphics[width=0.47\linewidth]{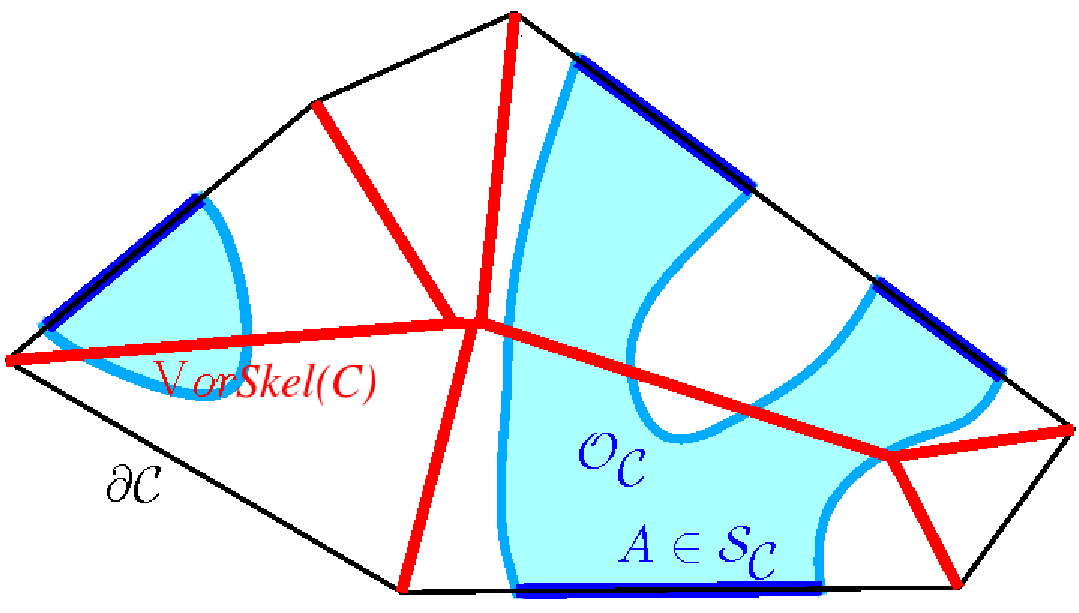}}
  \hspace{3mm}
 \subfigure{\includegraphics[width=0.47\linewidth]{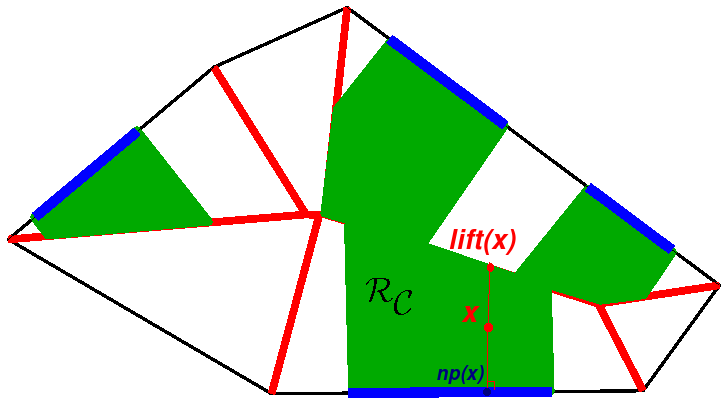}}
\caption{A 2D illustration of the partition of a cell $\C$ by the Voronoi Skeleton $\VD(\C)$ (in red). Left) The original shape $\OC$ (in blue). Right) The reconstructed object $\RC$ (in green).}
  \label{fig:definition}
  \end{center}
  \end{figure}

\paragraph*{\bf Voronoi diagram of a cell} For a face $f$ of $\C$, the Voronoi cell of $f$, denoted by $\vor_{\C}(f)$, is defined as the set of all points in $\C$ that have $f$ as the nearest face of $\C$, i.e.,
$$ \vor_{\C}(f) := \{ \ x \in \C \ | \ d(x,f)  \leq d(x,f') , \ \forall \ \mathrm{face} \ f' \ \mathrm{of}\ \C \ \}.$$
Here $d(.,.)$ stands for the Euclidean distance. The collection of all $\vor_{\C}(f)$ when $f$ runs over all the faces of $\C$ forms a partition of $\C$, called the {\em Voronoi diagram of $\C$}. Let $\VD(\C)$ denote the {\em Voronoi skeleton of } $\C$ which is the locus of all the points in $\C$ that are at the same distance from at least two faces of $\C$. See Figure~\ref{fig:definition}-left. 

\medskip

\paragraph*{\bf Nearest point characterization}
 For any point $x$ in $\C$, the {\em nearest point} in $\partial \C$ to $x$ is the orthogonal projection of $x$ onto the nearest face $f$ of $\C$. This projection is denoted by $\np_f(x)$.
The set of all nearest points to $x$ in $\partial \C$ is denoted by $\NP_\C(x)$. Note that for any $x \notin \VD(\C)$, $\NP_\C(x)$ is reduced to a single point. Based on this, and in order to simplify the presentation, sometimes we drop the index $f$, and by $\np(x)$ we denote a point of $\NP_\C(x)$. 


\begin{definition}[Lift function] {\rm Let $x \in \C$ be a point in the Voronoi cell of a face $f$ of $\C$. The {\it lift of $x$ in $\C$}, denoted by $\lift_\C(x)$ (or simply $\lift(x)$ if $\C$ is trivially implied), is defined to be the unique point of $\VD(\C)$ whose orthogonal projection onto $f$ is $\np(x)$, see Figure~\ref{fig:definition}-right.

\noindent --- The {\it lift of a set of points} $X \subseteq \C$, denoted by $\lift(X)$, is the set of all the points $\lift(x)$ for $x \in X$, i.e., $\lift(X):=\{\,\lift(x) \,|\, x\in X\,\}.$

\noindent --- The function $\LLL: \C \rightarrow \VD(\C)$ that maps each point $x \in \C$ to its lift in $\VD(\C)$ will be called the {\em lift function} in the sequel. For any $Y \subset \VD(\C)$, $\LLL^{-1}(Y)$ denotes the set of points $x \in \C$ such that $\lift(x)=y$ for some $y \in Y$.}
\end{definition}

We now present a geometric characterization of the reconstructed object using the described lifting procedure. 

\medskip

\paragraph*{\bf Characterization of the reconstructed object $\RC$} If $\SC =\emptyset$, then for any point $x \in \C$, $\np(x) \notin \SC$ and so $\RC$ is empty. Otherwise, let $A \in \SC$ be a section lying on a face of $\C$.  
For each point $a \in A$, the locus of all the points $x \in \C$ that have $a$ as their nearest point in $\partial \C$ is the line segment $[a,\lift(a)]$ joining $a$ to its lift. Therefore, the reconstructed object $\RC$ is the union of all the line-segments $[a,\lift(a)]$ for a point $a$ in a section $A \in \SC$, i.e.,
\[\RC := \ \{x \in \C | \np(x) \in \SC \} = \ \bigcup_{A \in \SC}\: \bigcup_{a \in A}\:\: [a,\lift(a)] = \ \LLL^{-1}(\LSC). \]

\noindent Note that according to this characterization, if the lifts of two sections intersect in $\VD(\C)$, then these two sections are connected in $\RC$. This generalizes the
classical overlapping criterion for the case of parallel cutting planes. The union of all the pieces $\RC$ over all cells $\C$ will be the overall reconstructed object $\RR$. 

The rest of the paper is devoted to prove that under two appropriate sampling conditions, $\RR$ and $\OO$ are homotopy equivalent, and are indeed homeomorphic (and isotopic). Let us first infer the following simple observation from the described characterization of $\RC$. 
\begin{proposition} 
\label{prop:RCLFS}
The lift function $\LLL: \RC \rightarrow \LSC$ is a homotopy equivalence.
\end{proposition}
 \begin{figure}[!htb]
   \begin{center}
 \subfigure{\includegraphics[width=0.42\linewidth]{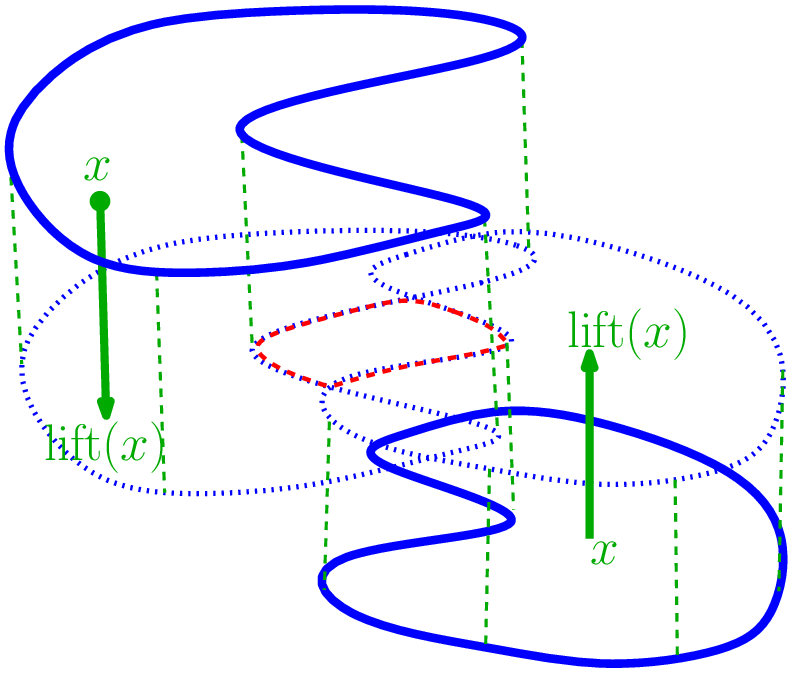}}
  \hspace{3mm}
 \subfigure{\includegraphics[width=0.42\linewidth]{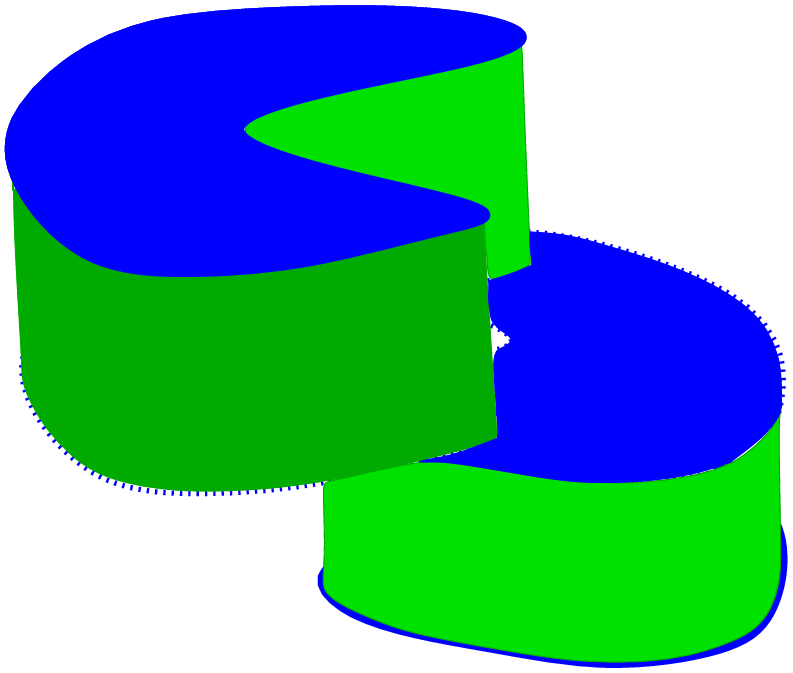}}
\caption{ A 3D reconstruction example from a pair of parallel sections (in blue). The lift function retracts $\RC$ (in green) onto the lift of the sections.}
  \label{retraction-fig}
  \end{center}
  \end{figure}

This is inferred trivially from the fact that the lift function retracts each segment $[a, \lift(a)]$ onto $\lift(a)$ continuously. See Figure~\ref{retraction-fig} for an example.

\subsection{First sampling condition: Separation Condition}
\label{sec:sampling}
In this section, we provide the first sampling condition, under which the connection between the sections in the reconstructed object $\RR$ will be the same as in the original object $\OO$. Our discussion will be essentially based on the study of the {\em medial axis} of $\partial \OO$, that we define now.

\begin{figure}[!htb]
    \begin{center}
      \subfigure{\includegraphics[width=0.47\linewidth]{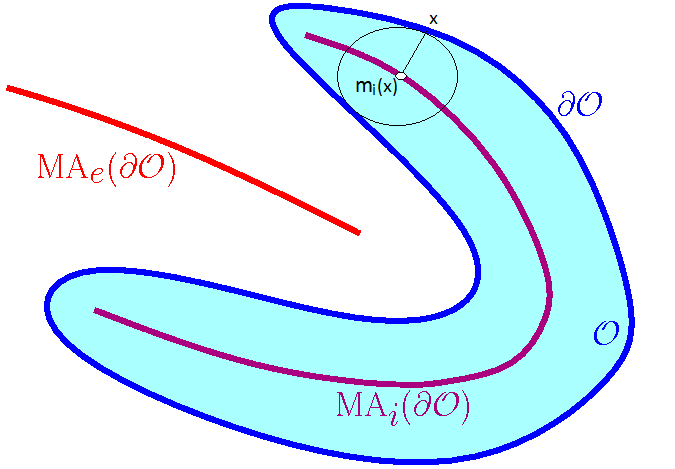}}
     \end{center}
 \caption{Internal and external parts of the medial axis of $\partial \OO$.}
 \label{ma-fig}
   \end{figure}

\begin{definition}[Internal and external parts of the medial axis of $\partial \OO$]\rm Consider $\partial \OO$ as a 2-manifold without boundary embedded in $\R^3$. The medial axis of $\partial \OO$ denoted by $\MA(\partial \OO)$
contains two different parts: the so-called {\em internal} part denoted by $\MA_i(\partial \OO)$, which lies in $\OO$, and 
the so-called {\em external} part denoted by $\MA_e(\partial \OO)$, which lies in $\R^3\setminus \OO$, see Figure~\ref{ma-fig}.

\end{definition}

The {\em internal retract} $m_i: \partial \OO \rightarrow \MA_i(\partial \OO)$ is defined as follows: for a point $x \in \partial \OO$, $m_i(x)$ is the
center of the maximum ball entirely included in $\OO$ which passes through $x$. For any  $x \in \partial \OO$, $m_i(x)$ is unique.
Symmetrically, we define {\em the external retract} $m_e: \partial \OO \rightarrow \MA_e(\partial \OO)$: for a point $x \in \partial \OO$, $m_e(x)$ is
the center of the maximum ball entirely included in $\R^3\setminus \OO$ which passes through $x$. For any  $x \in \partial \OO$, $m_e(x)$ is unique
but may be at infinity. In the sequel, we may write $m(a)$ for a point in $\{m_i(a),m_e(a)\}$. 

The interesting point is that as discussed below, if the set of cutting planes is sufficiently dense, then the internal part of $\MA(\partial \OO)$ lies inside the defined reconstructed object and the external part of this medial axis lies outside the reconstructed object. 


\begin{definition}[Separation Condition]
{\rm We say that the set of cutting planes verifies the Separation Condition if
$$\MA_i(\partial \OO) \subset \RR \ \ \mathrm{and} \ \ \MA_e(\partial \OO) \subset \R^3 \setminus \RR\,.$$
In other words, {\it $\partial \RR$ separates the internal and the external parts of the medial axis of $\partial \OO$.} (That is where the name comes from.)}
\end{definition}

\noindent We will show in Section~\ref{sec:ensure-separation} that the separation condition is ensured as soon as the set of cutting planes is sufficiently dense.

\medskip

\begin{figure}[!htb]
   \begin{center}
     \subfigure{\includegraphics[width=0.57\linewidth]{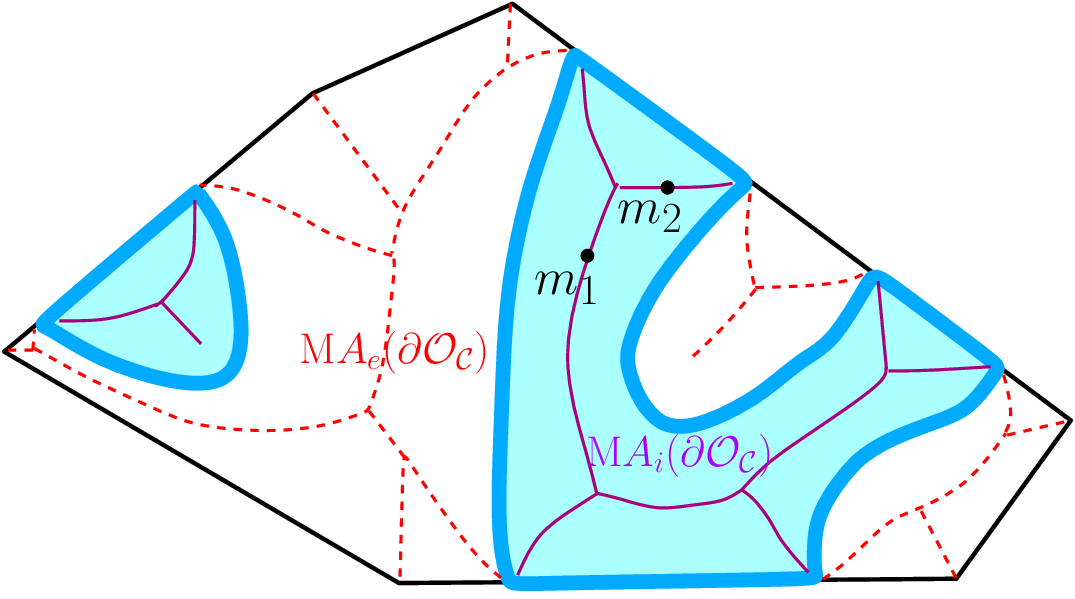}}
    \end{center}
\caption{2D example of medial axes in a cell $\C$ of the arrangement.}
\label{mac-fig}
  \end{figure}

\noindent {\bf Global versus local.} We now show that in each cell $\C$, the Separation Condition implies that $\partial \RC$ separates the internal and the external parts of the medial axis of $\partial \OC$. In order to study the Separation Condition in a cell $\C$, we will need the following definition: 

\begin{definition}[Medial axes in a cell $\C$ of the arrangement] \rm By $\MA_i(\partial \OC)$ we denote the set of all points in $\OC$ with at least two closest points in $\partial \OC$, see Figure~\ref{mac-fig}. Note that the two sets $\MA_i(\partial \OC)$ and $\MA_i(\partial \OO) \cap \C$ may be different. Symmetrically, $\MA_e(\partial \OC)$ denotes the medial axis of the closure of $\C \setminus \OC$. 
\end{definition}

\vspace{.3cm}

We also consider the {\em internal retract} $m_{i,\C}: \partial \OC \rightarrow \MA_i(\partial \OC)$ defined as follows. For a point $x \in \partial \OC$, $m_{i,\C}(x)$ is the
center of the maximum ball entirely included in $\OC$ which passes through $x$. Symmetrically, we can define {\em the external retract} $m_{e,\C}: \partial \OC \rightarrow \MA_e(\partial \OC)$: for a point $x \in \partial \OC$, $m_{e,\C}(x)$ is the center of the maximum ball entirely included in $\C \setminus \OC$ which passes through $x$. It is easy to see that for any $x \in \partial \OO \cap \C$, the segments $[x,m_{i,\C}(x)]$ and $[x,m_{e,\C}(x)]$ are subsegments of $[x,m_i(x)]$ and $[x,m_e(x)]$ respectively, and lie on the line defined by the normal to $\partial \OO$ at $x$.

\begin{lemma}[Separation Condition restricted to $\C$]
\label{scr}
{ If the Separation Condition is verified, then $\MA_i(\partial \OC) \subset \RC$ and $\MA_e(\partial \OC) \subset \C \setminus \RC$. 
}
\end{lemma}
\begin{proof}
We prove the first part, i.e., $\MA_i(\partial \OC) \subset \RC$. A similar proof gives the second part.
\noindent Let $m$ be a point in $\MA_i(\partial \OC)$. Let $B(m)$ be the open ball centered at $m$ which passes through the closest points to $m$ in $\partial \OC$. 
Two cases may happen:

\vspace{.3cm}
 $\bullet$ Either, the closest points to $m$ in $\partial \OC$ are in $\partial \OO$, see $m_1$ in Figure~\ref{mac-fig}. In this case $m$ is a point in $\MA_i(\partial \OO)$. The Separation Condition states that $\MA_i(\partial \OO) \subset \RR$, and so $m \in \RC= \RR \cap \C$.
 
 \vspace{.3cm}
 
$\bullet$ Otherwise, one of the closest points to $m$ in $\partial \OC$ is a point $a$ in some section $A \in \SC$, see $m_2$ in Figure~\ref{mac-fig}. If $a$ is on the boundary of $A$, then since along the section-contours $\partial \OC$ is non-smooth, $a$ lies in $\MA_i(\partial \OC)$ and coincides with $m$, and $m=a$ is trivially in $\RC$. Hence, we may assume that $a$ lies in the interior of  $A$. Therefore, the ball $B(m)$ is tangent to $A$ at $a$, and the line segment $[a,m]$ is orthogonal to $A$. Since $B(m) \cap \partial \C = \emptyset$, $m$ and $a$ are in the same Voronoi cell of the Voronoi diagram of $\C$. Thus, $a \in \SC$ is the nearest point in $\partial \C$ to $m$. By the definition of $\RC$, we deduce that $m \in \RC$. 
\end{proof}

\subsection{Guarantees on the connections between the sections}
\label{sec:connection}
We now show that if the set of cutting planes verifies the Separation Condition, then in each cell $\C$ of the arrangement, the connection between the sections is the same in $\OC$ and in $\RC$. 
\begin{theorem}
\label{P1}
{
If the set of cutting planes verifies the Separation Condition, 
$\RC$ and $\OC$ induce the same connectivity components on the sections of $\C$.
}
\end{theorem}

\begin{proof} We prove that two sections are connected in $\OC$ iff they are connected in $\RC$.
\begin{itemize}
\item[(I)] We first show that if two sections are connected in $\OC$, then they are connected in $\RC$.
Let $A$ and $A'$ be two sections in a same connected component $K$ of $\OC$. Due to the non-smoothness of $\partial \OC$ at the boundary of the sections, $\partial A$ and $\partial A'$ are contained in $\MA_i(\partial \OC)$ (more precisely in $\MA_i(\partial K)$). Thus, since $\MA_i(\partial K)$ is connected \cite{Lieutier}, there is a path $\gamma$ in $\MA_i(\partial K)$ that connects a point $a \in \partial A$ to a point $a' \in \partial A'$. According to Lemma~\ref{scr}, $\MA_i(\partial K) \subseteq \MA_i(\partial \OC) \subset \RC$. Thus, $\gamma$ is a path in $\RC$ that connects $A$ to $A'$. 

\item[(II)] To prove the other direction, 
let $A$ and $A'$ be two sections connected in $\RC$. We show that they are also connected in $\OC$. Let $\gamma$ be a path in $\RC$ that connects a point $a \in A$ to a point $a' \in A'$. For the sake of contradiction, suppose that $a$ and $a'$ are not in the same connected component of $\OC$. In this case, since $\gamma$ joins two points in two different connected components of $\OC$, it must intersect $\MA_e(\partial \OC)$. But this is a contradiction with the fact that $\gamma \subset \RC$, since according to Lemma~\ref{scr}, we have $\MA_e(\partial \OC) \cap \RC= \emptyset$.  
\end{itemize}
\end{proof}

\begin{figure}[!htb]
   \begin{center}
     \subfigure{\includegraphics[width=0.57\linewidth]{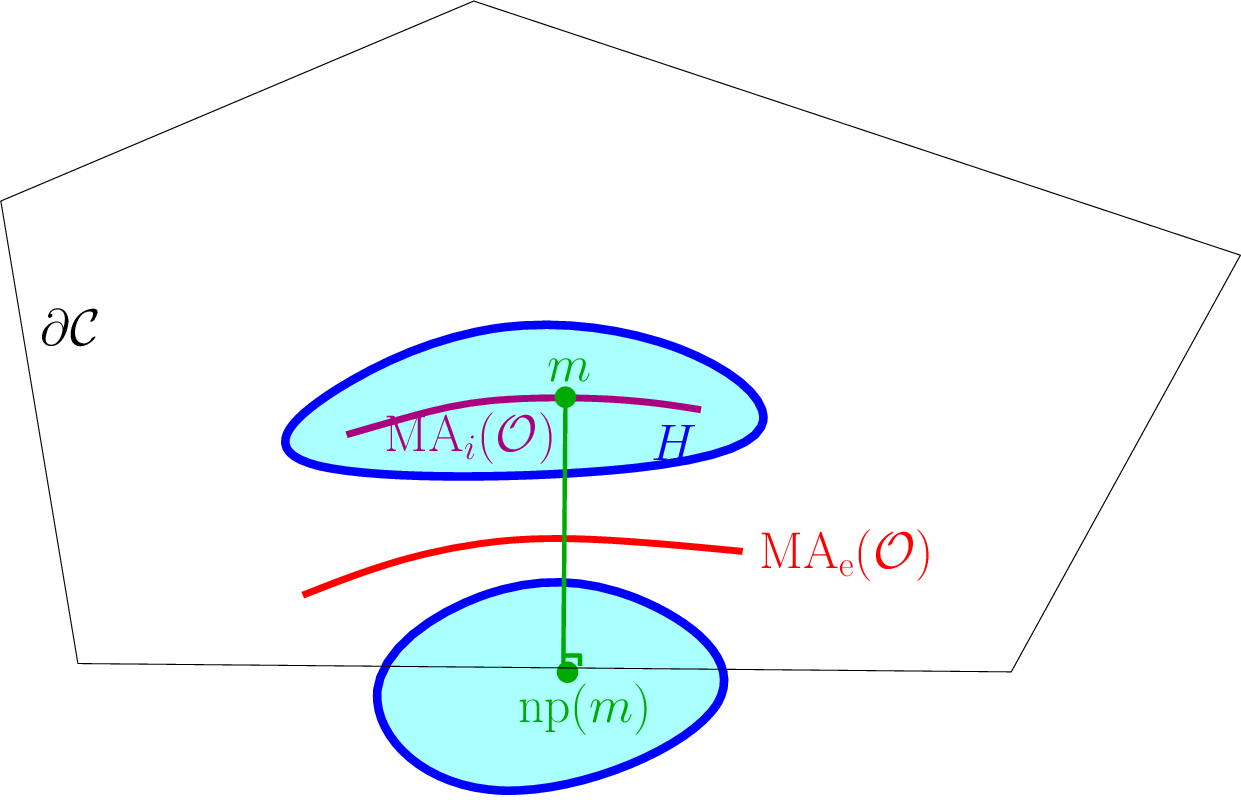}}
\caption{ For the proof of Proposition~\ref{cc}.} 
  \label{cc-fig}
  \end{center}
  \end{figure}

We state the following proposition that will be needed later in this section and in Section~\ref{chap:method1-homotopy}.
\begin{proposition}\label{cc}
\rm
Under the Separation Condition, any connected component of $\partial \OO$ is cut by at least one cutting plane.
\end{proposition}
\begin{proof}
\rm Suppose that $K$ is a connected component of $\partial \OO$ which is not cut by any cutting plane. There exists a cell $\C$ of the arrangement of hyperplanes such that one of the following two (symmetric) cases can happen: Either, there exists a connected component $H$ of $O$ which lies in the interior of $\C$ such that $K \subset \partial H$, Or, there exists a connected component $H$ of the closure of  $\mathbb R^3 \setminus \OO$ which lies in the interior of $\C$ such that $K \subset \partial H$. Without loss of generality, let us suppose the first case (the other case follows similarly). In this case, $\partial H$ bounds a connected component $H$ of $\OO$ in $\C$, $H$ is entirely contained in the interior of $\C$, and $K \subset \partial H$, see Figure~\ref{cc-fig}. Take a point $m$ in the medial axis of $H$, i.e., $m \in H \cap \MA_i(\partial \OO)$. According to the Separation Condition, $m$ belongs to $\RR$. Thus, by the definition of $\RR$, one of the nearest points of $m$ in $\partial \C$, say $\np(m)$, belongs to $\SSS$. Since $H$ is not cut by any cutting plane, $H \cap \partial \C$ is empty and $\np(m) \notin H$. Therefore, $m$ and $\np(m)$ are in two different connected components of $\OO$, and the segment $[m,\np(m)]$ should intersect $\MA_e(\partial \OO)$ at a point $x$. On the other hand, by the definition of $\RR$, the segment $[m,\np(m)] \subset \RR$. This contradicts the assumption of the Separation Condition that $\RR \cap \MA_e(\partial \OO) = \emptyset$.
\end{proof}

\subsection{How to ensure the Separation Condition?}
\label{sec:ensure-separation}
In this section we provide a sufficient condition for ensuring the Separation Condition. For this, we first need some definitions.

\begin{definition}[Reach] {\rm Let $\OO$ be a compact $3$-manifold with smooth boundary $\partial \OO$ in $\R^3$. For $a \in \partial \OO$, we define $\reach(a) :=\min \bigl(d(a, m_i(a)), d(a, m_e(a)) \bigr)$. 
The quantity $\reach(\OO)$ is defined as the minimum distance of $\partial \OO$ from the medial axis of $\partial \OO$:
$$\reach(\OO) \: :=  \min_{m \in \MA(\partial \OO)} \ d(m, \partial \OO)=  \min_{a \in \partial \OO} \reach(a).$$
Note that since $\OO$ is compact and $\partial \OO$ is of class $C^2$, 
$\reach(\OO)$ is strictly positive (see \cite{Fed59} for a proof).} 
\end{definition}
\begin{definition}[Reach restricted to a cell]
{\rm Given a cell $\C$ of the arrangement, 
we define $\reach_{\C}(\OO)  := \ \min d(a, m(a))$, where either $a$ or $m(a)$ is in $\C$. 
 By definition, we have $\reach(\OO)= \min_{\C} \ (\reach_{\C}(\OO))$.}
\end{definition}

\begin{figure}[!htb]
   \begin{center}
 \subfigure{\includegraphics[width=0.4\linewidth]{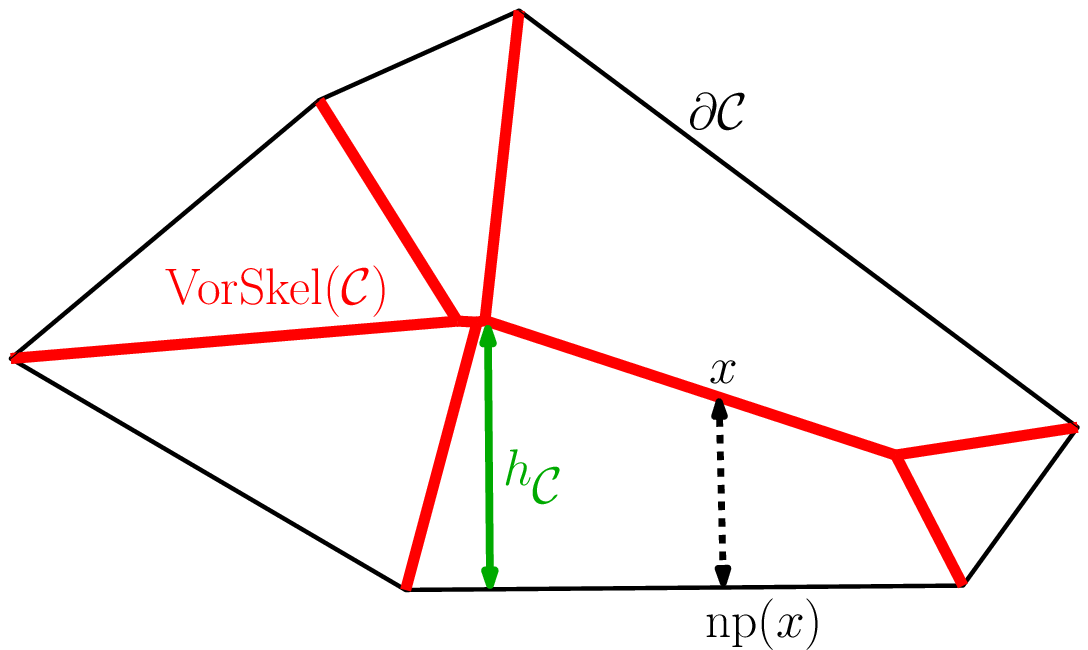}}
\hspace{2mm}
 \subfigure{\includegraphics[width=0.47\linewidth]{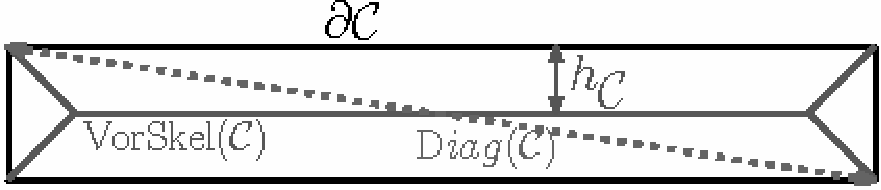}}
\caption{Left) Definition of the height of a cell $\C$. Right) Height of a cell $\C$ is bounded from above by half of the diameter of $\C$, but may be much smaller for many configurations of cutting planes.} 
  \label{hc-fig}
  \end{center}
  \end{figure}

\begin{definition}[Height of a cell]
{\rm
Let $\C$ be a cell of the arrangement of the cutting planes. The {\em height} of $\C$, denoted by $h_{\C}$, is defined as
the maximum distance of a point $x \in \C$ to its nearest point in the boundary of $\C$, see Figure~\ref{hc-fig}. In other words, $h_{\C} := \ \max_{x \in \C} \ d(x, \np(x)).$}
\end{definition}

We remark that the height of any cell $\C$ is at most half of the diameter of $\C$. However, as the example of Figure~\ref{hc-fig} (right figure) shows, it may be much smaller than half of the diameter. Moreover, in the case of parallel planes, while the cell between two consecutive planes is unbounded, the height of the cell is half of the distance between the two planes.
We now show that by bounding from above the height of the cells by the reach of the object, we can ensure the Separation Condition.

\begin{lemma}[Sufficient condition]
\label{lemma:ensure-separation}
{
If the set of cutting planes is sufficiently dense so that $h_{\C} < \reach_{\C}(\OO)$ for any cell $\C$ of the arrangement, then the Separation Condition is verified.
}
\end{lemma}
\begin{proof}
{
Let $m_i$ be any point in $\MA_i(\partial \OO)$ in a cell $\C$ of the arrangement. 
We have $$ d(m_i, \np(m_i)) \leq h_{\C} < \reach_{\C}(\OO) \leq d(m_i, \partial \OO).$$ Therefore, $\np(m_i)$ is in $\OO$ (and so in $\SSS$), and according to the definition of $\RR$, $m_i$ is in $\RR$. This proves that $\MA_i(\partial \OO) \subset \RR$. For any point $m_e \in \MA_e(\partial \OO)$, we can similarly show that $\np(m_e)$ is not in $\OO$, and so  $\MA_e(\partial \OO) \subset \R^3 \setminus \RR$. Therefore, the Separation Condition is verified.
}
\end{proof}

\begin{figure}[!htb]
   \begin{center}
\includegraphics[width=0.97\linewidth]{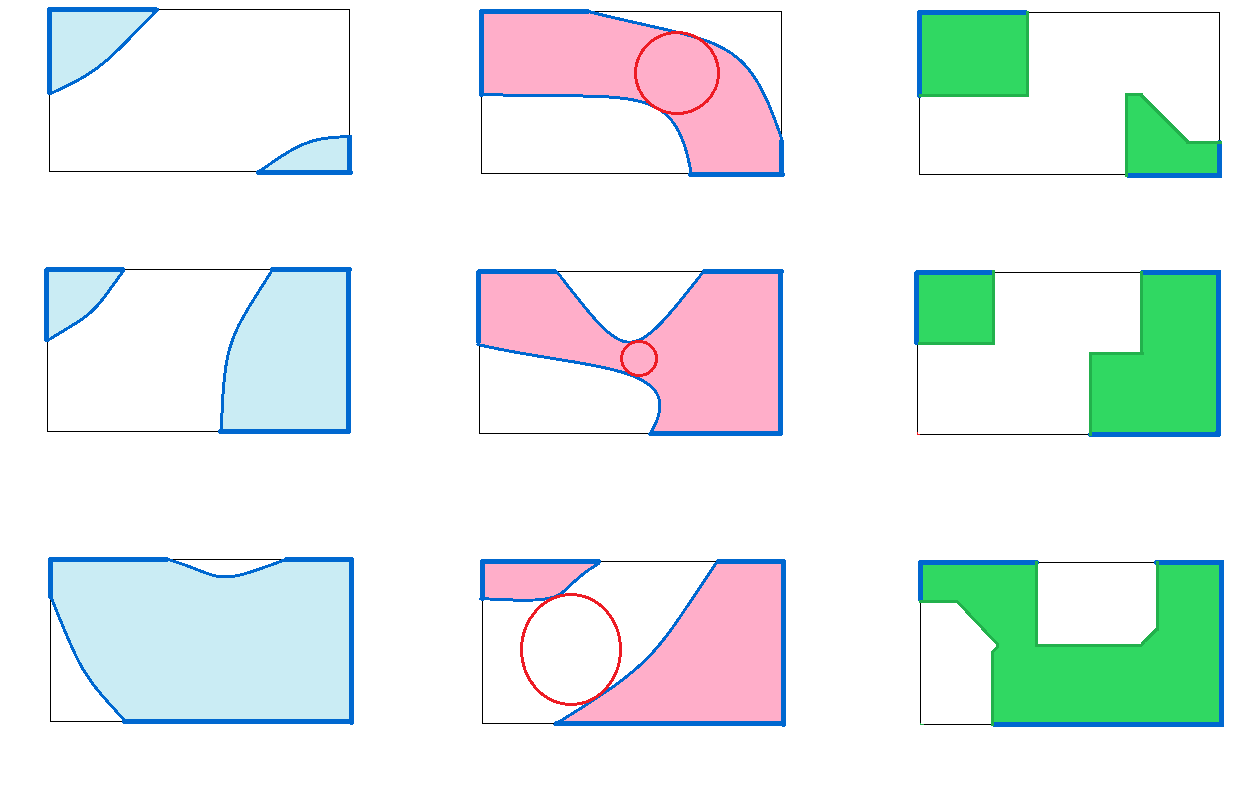}
\caption{The left column shows examples for which the sufficient condition is verified. The examples of the column in the middle have the same sections, but for each case there exists a medial ball with radius less than the height of the cell and so the condition is not verified. The right column shows the reconstructed object from the given sections for each case.} 
  \label{examples-fig}
  \end{center}
  \end{figure}

\begin{figure}[!htb]
   \begin{center}
 \subfigure[Original shape]{\includegraphics[width=0.42\linewidth]{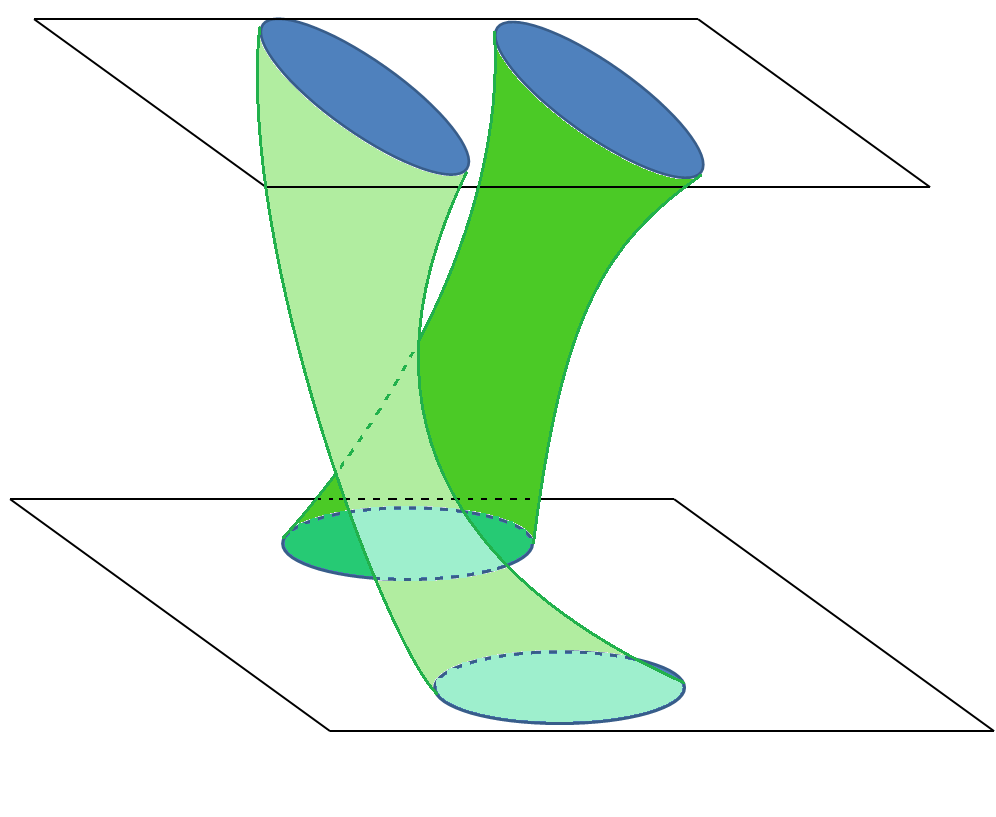}}
\hspace{2mm}
 \subfigure[Cross-sections]{\includegraphics[width=0.42\linewidth]{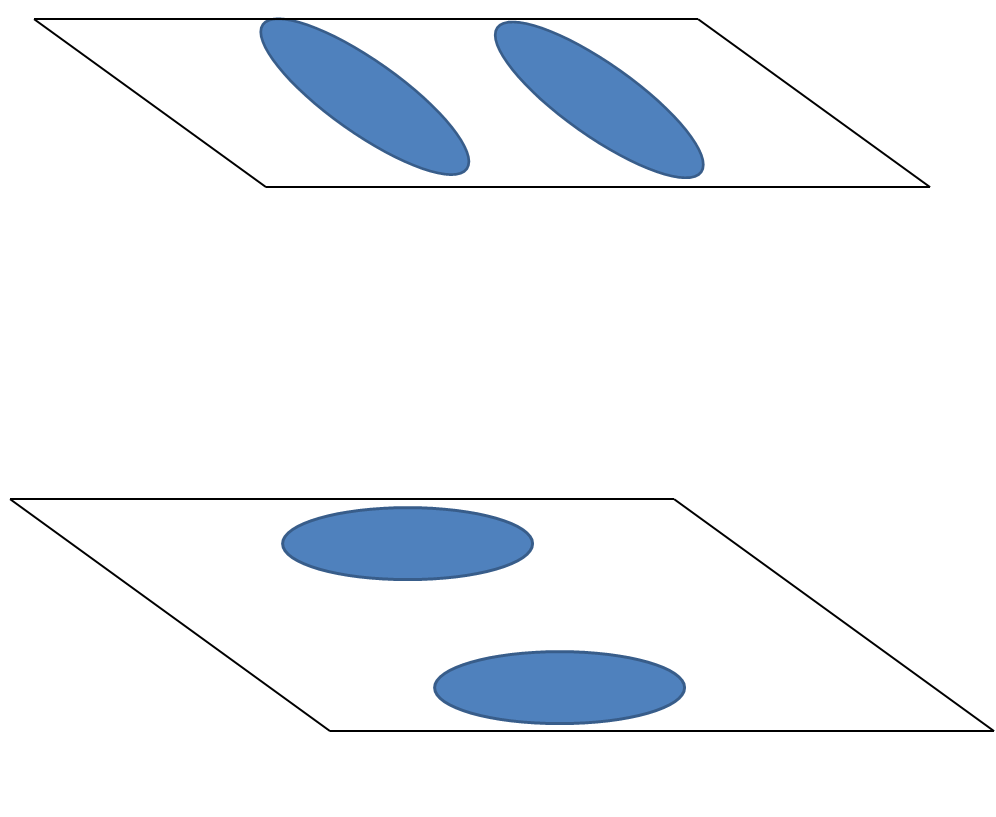}}
\hspace{2mm}
 \subfigure[Lift of the sections]{\includegraphics[width=0.42\linewidth]{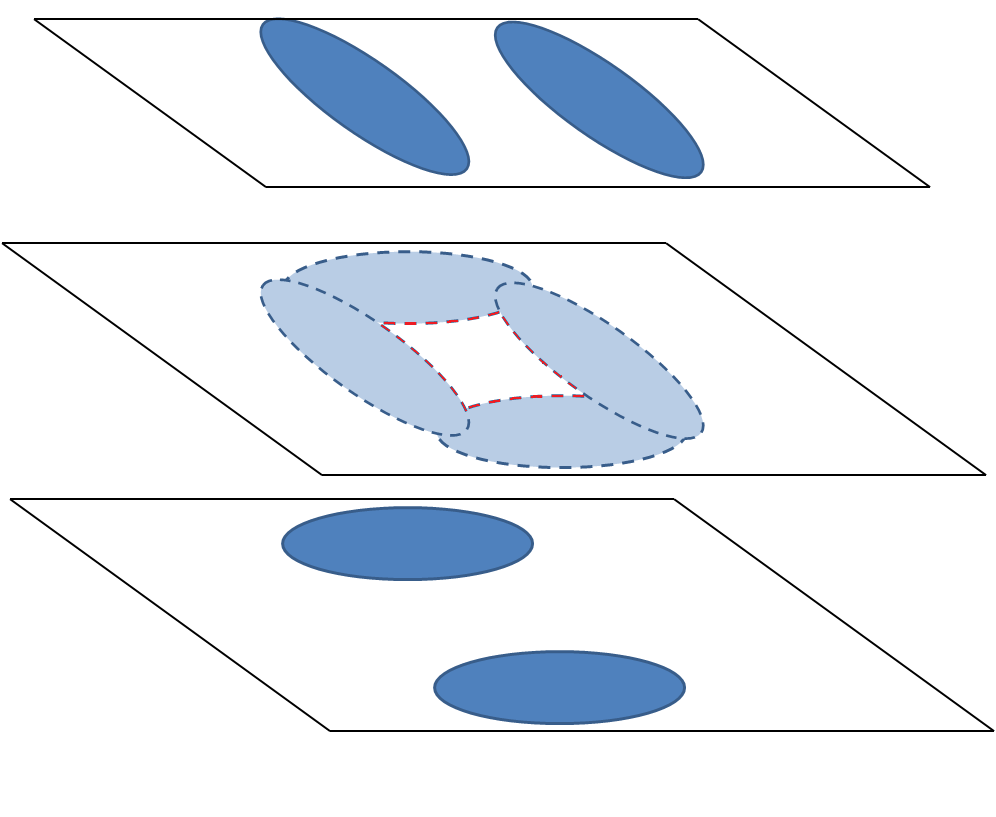}}
\hspace{2mm}
 \subfigure[Reconstructed object]{\includegraphics[width=0.42\linewidth]{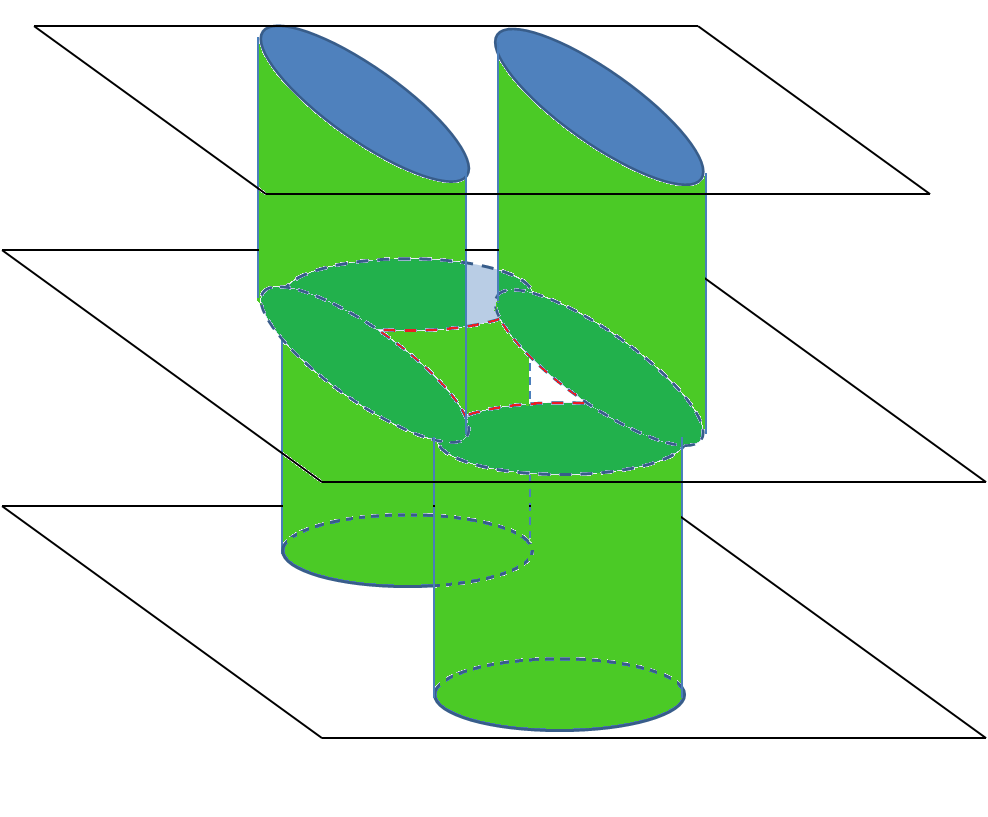}}
\caption{Example of bad sampling: the connectivity between the sections in the reconstructed object is not coherent with the original shape (two twisted cylinders). }
  \label{example-2-fig}
  \end{center}
  \end{figure}

Figure~\ref{examples-fig} shows some 2D examples in a rectangle cell. The height of the rectangle cell is obviously one half  the length of its smaller side. The left column shows examples for which the explained sufficient condition is verified. The examples of the column in the middle have the same sections, but for each case there exists a medial ball with radius less than the height of the cell and so the above condition is not verified. The right column shows the reconstructed object from the given sections of each case, inducing the same connectivity between the sections as the shapes of the left column. Figure~\ref{example-2-fig} illustrates another situation in 3D where the sampling of cutting planes is not sufficiently dense to verify the explained condition. 

\begin{figure}
   \begin{center}
   \includegraphics[width=0.85\linewidth]{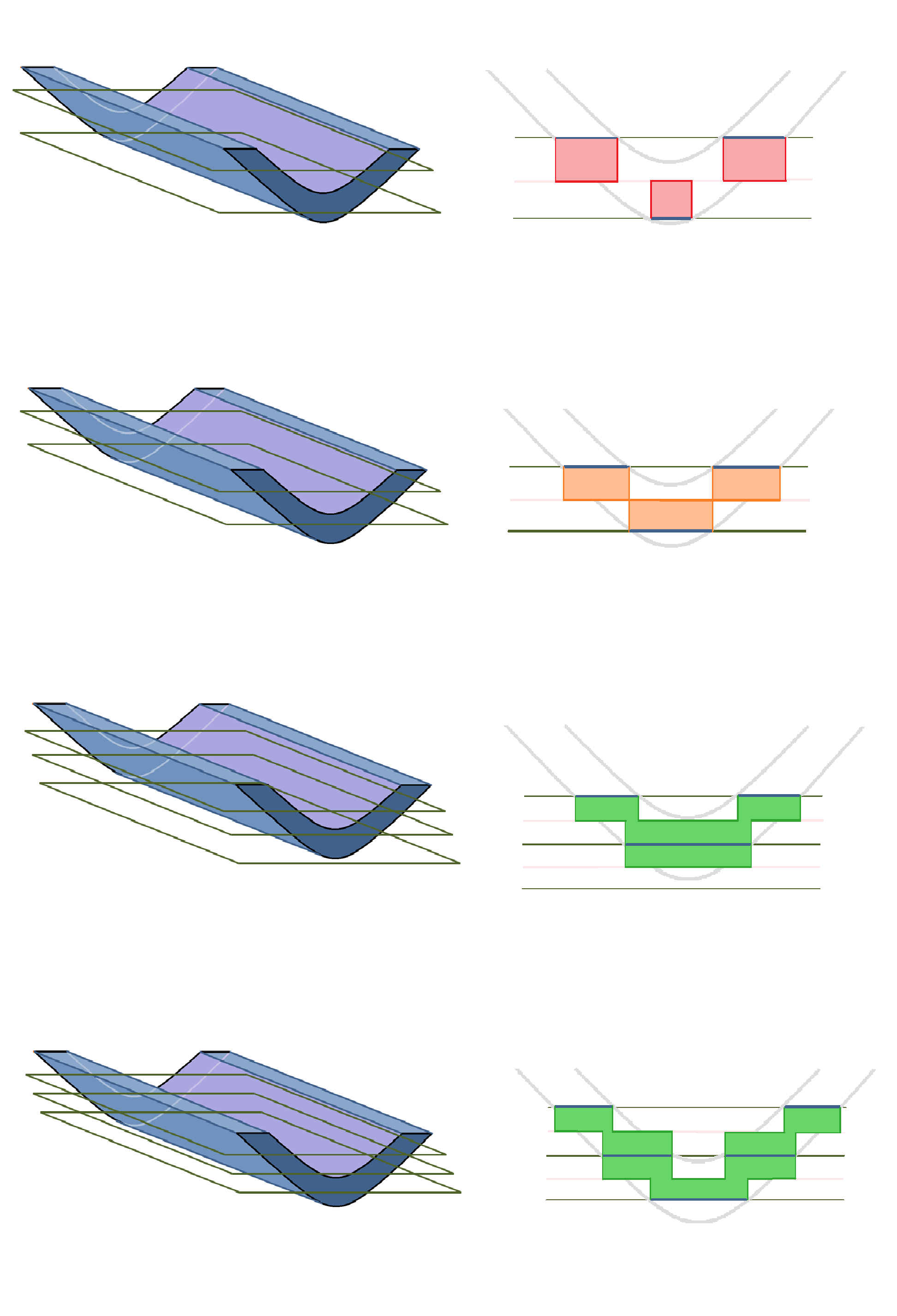}
\caption{Different sampling configurations. The distance between the cutting planes is respectively, (1) bigger than  (2) equal to  (3 and 4) less than twice the reach of the shape. Note that the front and rear cutting planes are not shown in the figures of the left column, only the corresponding sections are visible.} 
  \label{fig:3d}
  \end{center}
  \end{figure}

Figure~\ref{fig:3d} provides another series of examples for the case of parallel planes, for which the height of each cell is one half  the distance between the two adjacent planes. In the first example, the distance between the cutting planes is bigger than twice the reach of the shape. The previous sufficient condition is not verified and the connectivity is not correctly reconstructed. The second figure, illustrates two planes at distance exactly twice the reach of the shape. One can see how this critical value ensures the right connectivity. In the two last examples, the distance between cutting planes is small enough to ensure the Separation Condition, and thus, the correct connectivity between the sections.   

\section{Topological Guarantees for Particular Cases}
We proved that under the Separation Condition, the connectivity between the sections of $\C$ induced by the reconstructed object $\RC$ is coherent with the original shape $\OC$. We presented the proofs for the 3D case of the problem, but the proofs easily show that this result is valid in any dimension. We show in this section that this is strong enough to imply the homotopy equivalence between $\RC$ and $\OC$ for the 2-dimensional variant of the reconstruction problem, as well as for any disjoint union of convex bodies in $\R^d$. (In addition, as we will see later, the homotopy equivalence is also ensured for some other simple cases very common in practice, where the sections are contractible and sufficiently close to each other; more formally, if each connected component of the lift of the sections is contractible.) 
These guarantees justify the fact that under a rather simple density condition (i.e., Separation Condition), the overlapping criterion without any further information is enough to recover the topology for simple cases.

\subsection{Guarantees for 2D Shape Reconstruction from Line Cross-Sections}
\label{sec:2d}
Consider the 2-dimensional variant of the reconstruction problem, that consists of reconstructing a 2D-shape from its intersections with arbitrarily oriented {\em cutting lines}. 
In this case the sections are line-segments.

\noindent We can focus on a cell $\C$ of the arrangement of the plane by the cutting lines. Similar definitions for the Voronoi diagram and the Voronoi skeleton of $\C$, the lift function and the reconstructed object $\RC$ can be considered. 
Using the sufficient condition presented in the last section, if for any cell $\C$ of the arrangement, $h_{\C} < \reach_{\C}(\OO)$, then the Separation Condition is ensured. We deduce the following theorem.

\begin{theorem}[Provably good 2D reconstruction] 
If for any cell $\C$ of the arrangement of the cutting lines, $h_{\C} < \reach_{\C}(\OO)$, then $\RR$ is homeomorphic to $\OO$.
\end{theorem}

\begin{proof} 
By the definition of the reconstructed object, it is easy to see that any connected component of $\RC$ is a topological disk. On the other hand, according to Proposition~\ref{cc}, under the Separation Condition, any connected component of $\partial \OO$ is cut by at least one cutting line. We easily deduce that any connected component of $\OC$ is a topological disk. Therefore, all the connected components of $\OC$ or $\RC$ are 2-dimensional disks. 
On the other hand, according to Theorem~\ref{P1}, under the separation condition, there is a bijection between the connected components of $\RC$ and $\OC$. Therefore, there is a homotopy equivalence between each pair of corresponding connected components of $\OC$ or $\RC$. This provides a homotopy equivalence between $\RC$ and $\OC$. As we will explain in detail in Section~\ref{subsec:nerve}, the homotopy equivalences in the different cells of the arrangement can be extended to a homotopy equivalence between $\RR$ and $\OO$. Finally, since $\RR$ and $\OO$ are two homotopy equivalent 2-dimensional submanifolds of $\mathbb R^2$, we deduce that there is a homeomorphism between $\RR$ and $\OO$.
\end{proof}

\paragraph*{\bf Comparison with our results presented in \cite{MB08}} We note that in~\cite{MB08}, we presented a Delaunay-based 2-dimensional reconstruction algorithm from line cross-sections, and provided a sampling condition under which the same topological guarantees are ensured. Rather than using the Voronoi diagram of the cells, in \cite{MB08} the nearest point definition is based on the Voronoi diagram of all the sections, which performs well for the case of reconstructing tree-like objects from sparse sectional data.  
Note that since in that case the Voronoi cells are not restricted to the cells of the arrangement, the equivalent {\em height function} of the cells may be too large and the sampling condition may be more difficult to ensure. Therefore, the sampling condition provided in the present paper seems more appropriate.
 
\subsection{Reconstruction of a collection of disjoint convex bodies {\bf(arbitrary dimension)}}
\label{sec:convex}
Another interesting particular case is any disjoint  union of convex bodies in $\R^d$. In this case $\OC$, which is the intersection of a set of convex bodies with a convex polyhedron $\C$, is composed of convex components as well.  

We claim that the presented reconstruction method can be adapted to the case of a union of convex bodies so that the resulting reconstructed object is homotopy equivalent to $\OC$ under the Separation Condition. 

\medskip

\paragraph*{\bf Reconstruction algorithm adapted to a union of convex bodies} Consider the connectivity classes of the sections in $\RC$, i.e., groups of sections that are in the same connected component of $\RC$. We define the new reconstructed object $\mathrm{Conv}_\C$ as the union of the convex hulls of the connectivity classes of sections in $\RC$. In other words, for each group of sections $K$ that are in the same connected component of $\RC$, we consider the convex hull of all the sections of $K$, denoted by $\mathrm{conv}(K)$. The new reconstructed object is then defined as $\mathrm{Conv} := \bigcup_\C \mathrm{Conv}_\C$. 

\begin{lemma}\rm
Under the Separation Condition, the reconstructed object $\mathrm{Conv}$ conforms to the given sections $\SSS$, and is homotopy equivalent to $\OO$.
\end{lemma}
\begin{proof}
Consider any group $K$ of sections in $\SC$ that are in the same connected component of $\RC$. We first show that $K$ has at most one section in each face of $\C$. We use the fact that, under the Separation Condition, there is a bijection between the connected components of $\RC$ and $\OC$. According to this bijection, the sections of $K$ are in the same connected component of $\OC$ (and so of $\OO$). Therefore, all the sections of $K$ belong to a convex body in $\OO$. We deduce that each face of $\C$ intersects $K$ in at most one connected component (section). On the other hand, since $\C$ is convex, the convex hull of $K$ lies inside $\C$ and we have $\mathrm{conv}(K) \cap \partial \C = K$ for any $K$. 
Thus, for any cell $\C$, $\mathrm{Conv}_\C$ conforms to $\SC$. 

On the other hand, all the connected components of $\OC$ and $\mathrm{Conv}_\C$ are convex and form $d$-dimensional topological balls. Thus, the bijection between the connected components of $\OC$ and $\RC$ (and so $\mathrm{Conv}_\C$) provides a homotopy equivalence between the components that are all $d$-dimensional topological balls. Now the lemma follows by observing that $\mathrm{Conv}$ itself is a disjoint union of convex bodies, which are in bijection with the convex bodies in $\OO$. 
\end{proof}

\noindent Using the sufficient condition that implies the Separation Condition, c.f. Lemma~\ref{lemma:ensure-separation}, we deduce the following theorem.
 
\begin{theorem}[Reconstruction of a collection of convex bodies]
\rm Let $\OO$ be a union of convex bodies in $\R^d$. We define the new reconstructed object as $\mathrm{Conv} := \bigcup_\C \mathrm{Conv}_\C$, where $\mathrm{Conv}_\C$ is the union of the convex hulls of the connectivity classes of sections in $\RC$. If for any cell $\C$ of the arrangement of the cutting lines, $h_{\C} < \reach_{\C}(\OO)$, then 
 $\mathrm{Conv}$ conforms to the given sections $\SSS$, and is homotopy equivalent to $\OO$.
\end{theorem}

We showed that the Separation Condition implies that the connectivity between the sections is correctly reconstructed. However, it only implies the homotopy equivalence between the reconstructed object and the original shape for the 2D variant of the problem or some particular cases as the case of a union of convex bodies. In the next section, we will impose a second sampling condition on the set of cutting planes in order to ensure the homotopy equivalence in the general 3D case.
\section{General Topological Guarantees}
\label{chap:method1-homotopy}
To clarify the connection between the upcoming sections, let us shortly outline the general strategy employed in proving the homotopy equivalence between $\RR$ and $\OO$.

\subsection{Proof outline of the homotopy equivalence between $\RR$ and $\OO$}
\label{sec:proofoutline}

In Section~\ref{sec:connection} we showed that under the first sampling condition called the Separation Condition, the connection between the sections in the reconstructed object $\RC$ is the same as in $\OC$, in the sense that there is a bijection between the connected components of $\RC$ and the connected components of $\OC$, for any cell $\mathcal{C}$.  This implies that for proving the homotopy equivalence between $\RC$ and $\OC$, it will be enough to show that the corresponding connected components have the same homotopy type. 
\begin{quote}
{\it In the sequel,  to simplify the notations and the presentation, we suppose that $\OC$ and thus $\RC$ are connected, and we show that $\OC$ and $\RC$ have the same homotopy type. It is clear that the same proofs can be applied to each corresponding connected components of $\OC$ and $\RC$ to imply the homotopy equivalence in the general case of multiple connected components. }
\end{quote}
The idea consists in following the normal-directions to $\partial \OO$ in order to define a deformation retract of $\OC$ onto a subset of $\RC$, in each cell $\C$. To this end, we will define an intermediate shape $\MC$, called the {\em medial shape}, such that $\MC \subset \RC$ and the map $\OC \rightarrow \MC \hookrightarrow \RC$, obtained by composing the deformation retract and the inclusion, restricts to the identity map on each section of $\SC$. Thus, we can glue all these maps to obtain a global map from $\OO$ to $\RR$. 

Using a generalized version of the nerve theorem (see Section~\ref{subsec:nerve}), we then reduce the problem to proving that 
the inclusion $i: \MC \hookrightarrow \RC$ forms a homotopy equivalence in each cell. Using Whitehead's theorem, it will be enough to show that the inclusion $i$ induces isomorphisms between the corresponding homotopy groups.  Under the Separation Condition, we prove that $i$ induces an injective map on the first homotopy groups, and that all higher dimensional homotopy groups of $\MC$ and $\RC$ are trivial. Unfortunately, the Separation Condition does not ensure in general  the surjectivity of $i$ on the first homotopy groups. To overcome this problem, we need to impose a second condition called {\it Intersection Condition}, under which the map $i$ will be surjective on the first homotopy groups, leading to a homotopy equivalence between $\OO$ and $\RR$.

\subsection{Medial shape}
\label{sec:MS}
\begin{figure}[!htb]
   \begin{center}
     \subfigure{\includegraphics[width=0.47\linewidth]{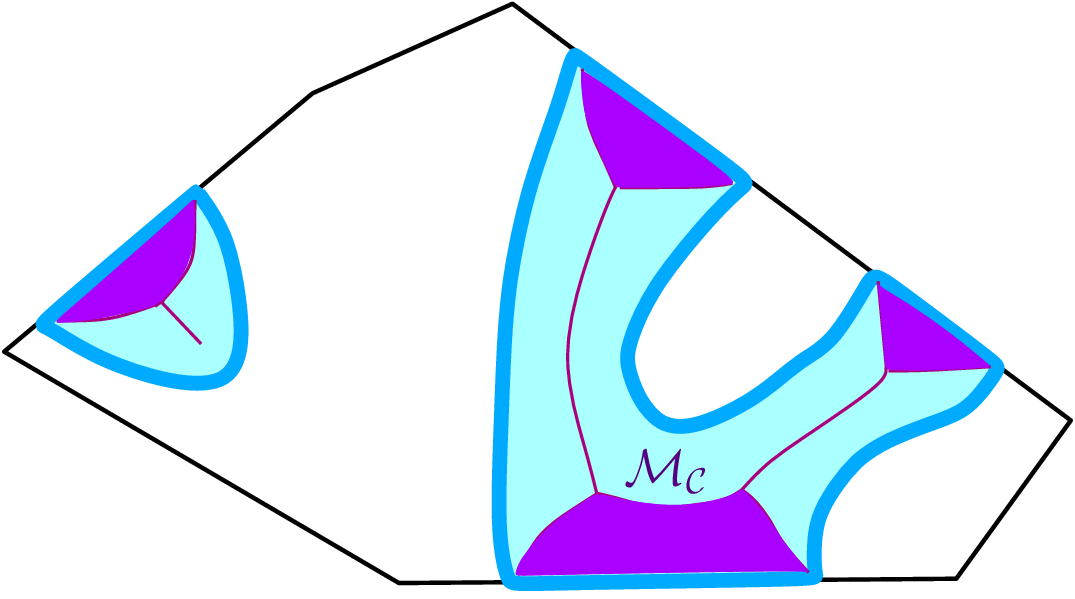}}
     \subfigure{\includegraphics[width=0.47\linewidth]{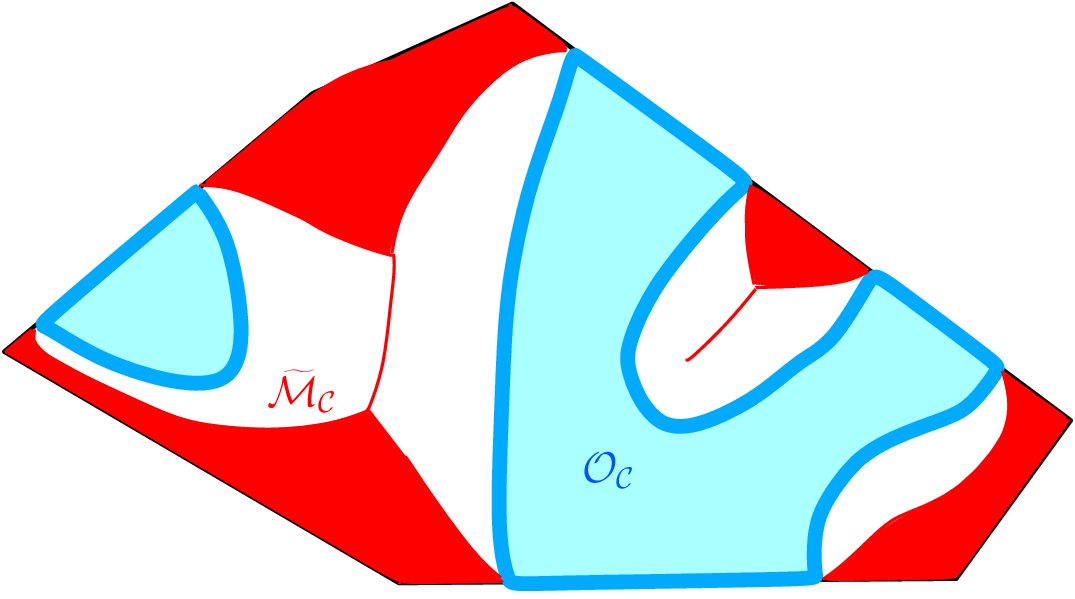}}
\caption{ A 2D illustration of the medial shape $\MC$ in purple (left) and $\WMC$ in red (right).} 
  \label{fig:MC}
  \end{center}
  \end{figure}

\begin{definition}[Medial shape $\MC$]
{\rm Let $x$ be a point in $\SC \subset \partial \OC$. Let $w(x)=[x,m_{i,\C}(x)]$ be the segment in the direction of the normal to $\partial \OC$ at $x$ which connects $x$ to the point $m_{i,\C}(x) \in \MA_i(\partial \OC)$. We add to $\MA_i(\partial \OC)$ all the segments $w(x)$ for all the points $x \in \SC$. We call the resulting shape $\MC$, see Figure~\ref{fig:MC} (left) for an example. 
More precisely, $$\MC := \MA_i(\partial \OC) \cup (\bigcup_{x\in \SC} w(x)).$$ }
\end{definition}
\begin{proposition}\label{prop:deform}
{ 
The medial shape verifies the following set of properties:
\begin{itemize}
\item[$(i)$] The medial shape contains the sections of $\C$, i.e., $\SC \subseteq \MC$. 
\item[$(ii)$] There is a (strong) deformation retract $r$ from $\OC$ to $\MC$. In particular, this map is a homotopy equivalence between $\OC$ and $\MC$. And its restriction to $\SC$ is the identity map.

\item[$(iii)$] Under the Separation Condition, $\MC \subseteq \RC$. 
\end{itemize}
}
\end{proposition}
\begin{proof} 
\begin{itemize}
\item[$(i)$] This property is true by the definition of the medial shape. 
\item[$(ii)$] This is obtained by deforming $\OC$ to $\MC$ in the direction of the normals to the boundary $\partial \OC$. Note that the boundary $\partial \OC$ is smooth except on the boundaries of sections in $\SC$, and the boundaries of the sections in $\SC$ 
 are already in $\MC$, thus, the deformation retract is well defined.
Moreover, since $\partial \OO$ (and so $\partial \OO \cap \mathcal{C}$) is supposed to be of class $C^2$, according to a theorem by Wolter (\cite{Wol92}), this deformation is continuous and is hence a continuous deformation retract from $\OC$ to $\MC$. 

\item[$(iii)$] Since $\MC= \MA_i(\partial \OC) \cup \bigl(\bigcup_{x \in \SC} w(x)\bigr)$ and in addition $\MA_i(\partial \OC) \subset \RC$, it will be sufficient to show that for any $x$ in a section $A \in \SC$, $w(x) \subset \RC$. (Recall that $w(x)$ is the orthogonal segment to $\partial \OC$ at $x$  that joins $x$ to the corresponding medial point $m_{i,\C}(x)$ in $\MA_i(\partial \OC)$.)   We will show that $w(x)$ is contained in the segment $[x, \lift(x)]$.
The point $x$ is the closest point in $\partial \OC$ to $m_{i,\C}(x)$. Thus, the ball centered at $m_{i,\C}(x)$ and passing through $x$ is entirely contained in $\OO$ and its interior is empty of points of $\partial \C$. Thus, in the Voronoi diagram of $\C$, $m_{i,\C}(x)$ is in the same Voronoi cell as $x$.  On the other hand, $x$ is the closest point in $\SC \subset \partial \OC$ to $\lift(x)$. It easily follows that $d(x,\lift(x)) \geq d(x, m_{i,\C}(x))$. 
It follows that the segment $[x, m_{i,\C}(x)]=w(x)$ is a subsegment of $[x, \lift(x)]$. Therefore, by the definition of $\RC$, $w(x) \subset \RC$.
\end{itemize}
\end{proof}

We end this section with the following important remark and proposition which will be used in the next section. By replacing the shape $\OC$ with its complementary set we may define an {\it exterior medial shape} $\WMC$. This is more precisely defined as follows.  Let $\widetilde \OO$ be the closure of the complementary set of $\OO$ in $\R^3$. 
And let $\widetilde \OC$ be the intersection of $\widetilde\OO$ with the cell $\C$. The medial shape of $\widetilde \OC$, denoted by $\WMC$, is the union of the medial shapes of the connected components of $\widetilde \OC$, see Figure~\ref{fig:MC} (right). Similarly, under the Separation Condition, the following proposition holds.
\begin{proposition}
\label{prop:deformrev}
 Let $\widetilde\OC$ be the closure of the complementary set of $\OC$ in $\C$ and $\WMC$ be the medial shape of $\widetilde \OC$. Under the Separation Condition: $(i)$ There is a strong deformation retract from $\C \setminus \WMC $ to $\OC$, and $(ii)$ We have $\RC \subset \C \setminus \WMC.$
\end{proposition}

\begin{proof}The proof of Property $(i)$ is similar to the proof of Proposition~\ref{prop:deform} by deforming along the normal vectors to the boundary of $\widetilde\OC$. The property $(ii)$ is equivalent to $\WMC \subset \C \setminus \RC$.
\end{proof}

\subsection{Topological guarantees implied by the Separation Condition}
\label{sec:homotopy}
Throughout this section, we suppose that the Separation Condition holds. By the discussion at the end of Section~\ref{sec:proofoutline}, and without loss of generality, we may assume that $\OC$ and hence $\RC$ are connected. 
Thus, $\OC$ and $\RC$ are connected compact topological 3-manifolds embedded in $\R^3$.

We showed that the medial shape $\MC$ is homotopy equivalent to $\OC$, and under the Separation Condition, $\MC \subset \RC$. Using these properties, we will show that $\LLL: \MC \rightarrow \LSC$ (the restriction of the lift function to $\MC$) is a homotopy equivalence. On the other hand, according to Proposition~\ref{prop:RCLFS}, $\LLL: \RC \rightarrow \LSC$ is a homotopy equivalence as well. Hence, using the following commutative diagram, we can infer that $i:\MC \hookrightarrow \RC$ is a homotopy equivalence.
\[
   \begin{diagram}
     \node{\MC} \arrow{e,t,J}{i} \arrow{se,r}{\mathcal L} 
       \node{\RC}  \arrow{s,r}{\mathcal L} \\
     \node[2]{\textrm{lift}(\SC)}
   \end{diagram}
\]
Moreover, since the objects we are manipulating are all CW-complexes, according to Whitehead's theorem homotopy equivalence is equivalent to weak homotopy equivalence. Hence, it will be enough to show that $\LLL: \MC \rightarrow \LSC$ induces isomorphism between the corresponding homotopy groups.

\subsubsection*{Injectivity at the level of homotopy groups}
We first show that under the Separation Condition, $\LLL: \MC \rightarrow \LSC$ induces injections at the level of homotopy groups.

\begin{theorem}[Injectivity]

Under the Separation Condition, the homomorphisms between the homotopy groups of $\MC$ and $\LSC$, induced by the lift function $\LLL$, are injective. 
\end{theorem}

\begin{proof}
 Under the Separation Condition, we have $\MC \subset \RC$. Let $\widetilde{ \MC}$ be the medial shape of the closure of the complementary set of $\OC$ in $\C$. We refer to the discussion at the end of the previous section for more details. Recall that by Proposition~\ref{prop:deformrev}, we have $\RC \subset \C \setminus \WMC$, and there exists a deformation retract from $\C \setminus \WMC$ to $\OC$ (in particular $\OC$ and $\C \setminus \WMC$ are homotopy equivalent).  We have now the following commutative diagram in which every map (except the lift function $\LLL$) is an injection (or an isomorphism) at the level of homotopy groups.
\[
 \begin{diagram}
 \node[3]{\OC} \arrow{sww,l}{\simeq}\\
     \node{\MC} \arrow[2]{e,t,J}{i} \arrow{see,b}{\mathcal L}
       \node[2]{\RC} \arrow[2]{e,J}\arrow{s,r}{\simeq}
         \node[2]{\C \setminus \WMC} \arrow{nww,t}{\simeq} \\
     \node[3]{\textrm{lift}(\SC)}  
   \end{diagram}
\]

Using this diagram, the injectivity at the level of homotopy groups is clear: For any integer $j \geq 1$, consider the induced homomorphism $\LLL_*: \pi_j(\MC) \rightarrow \pi_j(\LSC)$. Let $x \in \pi_j(\MC)$ be so that $\LLL_*(x)$ is the identity element of $\pi_j(\LSC)$. It is sufficient to show that $x$ is the identity element of $\pi_j(\MC)$. Following the maps of the diagram, and using the homotopy equivalence between $\LSC$ and $\RC$, we have that $x$ is first mapped to the identity element of $\pi_j(\RC)$ and then, by the inclusion $\RC \hookrightarrow \C \setminus \WMC$, it goes to the identity element of $ \C \setminus \WMC$. Finally, by the two retractions, it will be mapped to the identity element of $\pi_j(\MC)$. As this diagram is commutative, we infer that $x$ is the identity element of $\pi_j(\MC)$.
Thus, $\LLL_*: \pi_j(\MC) \rightarrow \pi_j(\LSC)$ is injective for all $j \geq 1$. The injectivity for $j=0$ is already proved in Theorem~\ref{P1}.
\end{proof} 

We have shown that under the Separation Condition, the lift function $\LLL: \MC \rightarrow \LSC$ induces injective morphisms between the homotopy groups of $\MC$ and $\LSC$. If these induced morphisms were surjective, then $\LLL$ would be a homotopy equivalence (by Whitehead's theorem). 
We will show below that the Separation Condition implies the surjectivity for all the homotopy groups except for dimension one (fundamental groups). Indeed, we will show that under the Separation Condition, all the $i$-dimensional homotopy groups of $\MC$ and $\LSC$ for $i \geq 2$ are trivial. Once this is proved, it will be sufficient to study the surjectivity of $\LLL_*: \pi_1(\MC) \rightarrow \pi_1(\LSC)$.

\begin{remark}\rm Note that the injectivity in the general form above remains valid for the corresponding reconstruction problems in dimensions greater than three. However, the vanishing results on higher homotopy groups of $\OC$ and $\RC$ are only valid in dimensions two and three.  
\end{remark}
\subsubsection*{The topological structures of $\RC$ and $\OC$ are determined by their fundamental groups}

We now show that if the Separation Condition is verified, then the topological structure of the portion of $\OO$ in a cell $\C$ (i.e., $\OC$) is simple enough, in the sense that for all $i \geq 2$, the $i$-dimensional homotopy group of $\OC$ is trivial. We can easily show that $\RC$ has the same property.\footnote{Recall that for simplifying the presentation, we assume that $\OC$ and so $\RC$ are connected. The same proof shows that in the general case, the same property holds for each connected component of $\OC$ or $\RC$.} 
As a consequence, the topological structures of $\OC$ and $\RC$ are determined by their fundamental group, $\pi_1(\OC)$ and $\pi_1(\RC)$. 

We first state the following general theorem for an arbitrary embedded $3$-manifold with connected boundary. 
\begin{theorem}
\label{Hi}
  Let $K$ be a connected 3-manifold  in $\mathbb R^3$ with  a (non-empty) connected boundary. Then for all $i \geq 2$, $\pi_i(K)= \{0\}$.
\end{theorem}

This theorem can be easily obtained from the Sphere Theorem, see e.g., Corollary 3.9 of \cite{Hatcher} or for more details~\cite{PooranThesis}. 
From this theorem, we infer the two following theorems.
\begin{theorem}
\label{th:OC}
 Under the Separation Condition, $\pi_i(\OC)= \{0\}$, for all $i \geq 2$.
\end{theorem}
\begin{proof} 
We only make use of the fact that under the Separation Condition, any connected component of $\partial \OO$ is cut by at least one cutting plane (Proposition~\ref{cc}). In this case, every connected component of $\OC$ is a 3-manifold with connected boundary. The theorem follows as a corollary of Theorem~\ref{Hi}.
\end{proof}
\begin{theorem}
\label{th:RC}
 $\pi_i(\RC)= \{0\}$, for all $i \geq 2$.
\end{theorem}
\begin{proof} Using Theorem~\ref{Hi}, it will be sufficient to show that the boundary of any connected component $K$ of $\RC$ is connected. Let $x$ and $y$ be two points on the boundary of $K$, and let $S$ and $S'$ be two sections so that $x \in [a,\lift(a)]$ for some $a \in S$ and $y \in [b,\lift(b)]$ for some $b \in S'$. By the definition of $\RC$, $x$ is connected to $S$ in $\partial \RC$, and $y$ is connected to $S'$ in $\partial \RC$. On the other hand, since $S$ and $S'$ are two sections in the connected component $K$ of $\RC$, they lie on $\partial K$ and are connected to each other in $\partial K$ (and so in $\partial \RC$). Thus, $x$ is connected to $y$ in  $\partial \RC$. 
\end{proof}

\subsection{Second condition: Intersection Condition}
\label{sec:intersection-condition}

In the previous section, we saw that under the Separation Condition, the topological structures of $\OC$ and $\RC$ are determined by their fundamental group $\pi_1(\OC)$ and $\pi_1(\RC)$, respectively. The goal of this section is to find a way to ensure an isomorphism between the fundamental groups of $\RC$ and $\OC$. We recall that as $\OC$ and $\MC$ are homotopy equivalent, $\pi_1(\OC)$ is isomorphic to $\pi_1(\MC)$. On the other hand, $\RC$ and $\LSC$ are homotopy equivalent, and $\pi_1(\RC)$ is isomorphic to $\pi_1(\LSC)$ (see last diagram). Thus, it will be sufficient to compare $\pi_1(\MC)$ and $\pi_1(\LSC)$.

\vspace{.3cm}
\noindent We consider $\LLL_*: \pi_1(\MC) \rightarrow \pi_1(\LSC)$, the map induced by the lift function from $\MC$ to $\LSC$ on fundamental groups. We showed that $\LLL_*$ is injective. 
A simple sufficient condition to ensure that $\LLL_*$ is an isomorphism is to impose that {\em $\LSC$ is contractible} (or more generally, each connected component of $\LSC$ is contractible). This is very common in practice, where the sections are contractible and sufficiently close to each other. In this case, all the homotopy groups $\pi_j(\LSC)$ are trivial and by injectivity of $\LLL_*$ proved in the previous section, $\LLL_*$ becomes an isomorphism in each dimension. Hence, the homotopy equivalence between $\RC$ and $\OC$ can be deduced.

 \begin{figure}[!htb]
   \begin{center}
\subfigure{\includegraphics[width=0.37\linewidth]{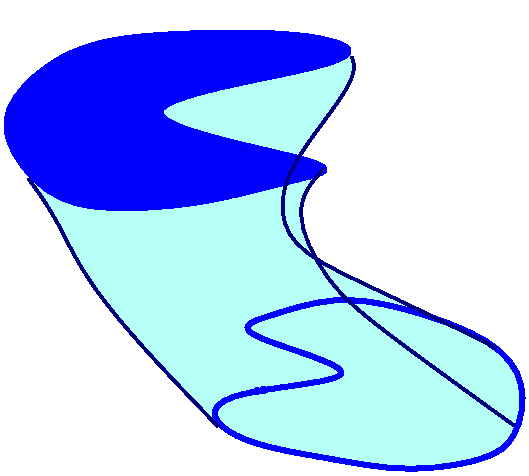}}
\hspace{.42mm}
\subfigure{\includegraphics[width=0.37\linewidth]{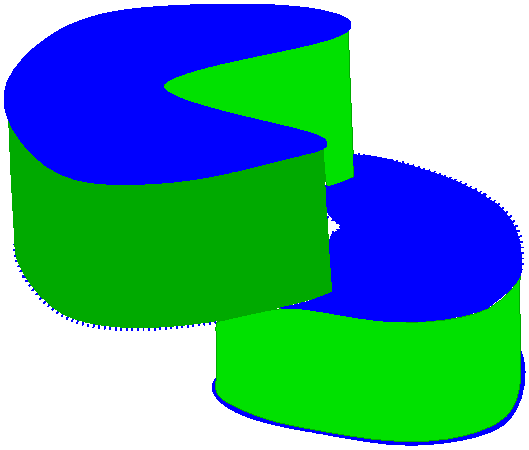}} \\
\subfigure{\includegraphics[width=0.37\linewidth]{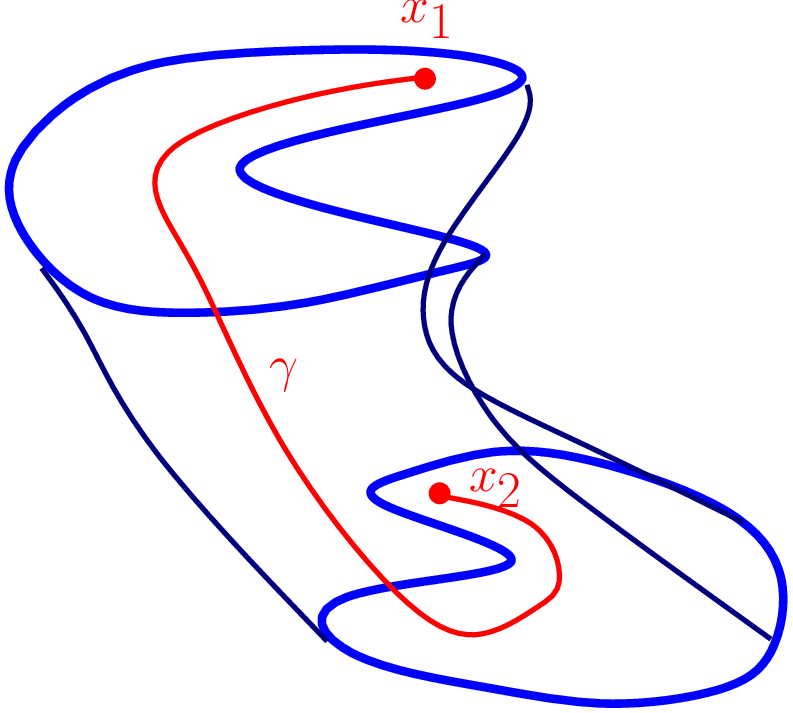}}
\hspace{.42mm}
\subfigure{\includegraphics[width=0.37\linewidth]{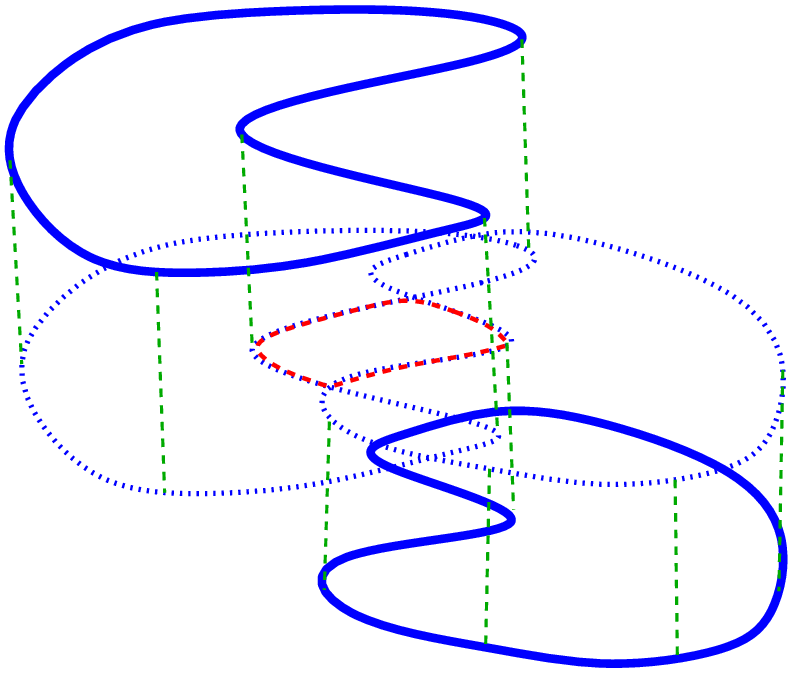}}
  \caption{ Left) Original shape. Right) Reconstructed shape. 3D example of the case where the lift function from $\MC$ to $\LSC$ fails to be surjective on the first homotopy groups.  Note that for two points  $x_1$ and $x_2$ with the same lift in $\LSC$, the lift of any curve $\gamma$ connecting $x_1$ and $x_2$ in $\MC$ provides a non-zero element of $\pi_1(\LSC,x)$. The reconstructed shape is a torus and is not homotopy equivalent to the original shape which is a twisted cylinder. } 
  \label{fig:torus-cylinder}
  \end{center}
  \end{figure}

However, the connected components of $\LSC$ are not necessarily contractible in general and the map $\LLL_*: \pi_1(\MC) \rightarrow \pi_1(\LSC)$ may fail to be surjective. Figure~\ref{fig:torus-cylinder} shows two shapes with different topologies, a torus and a ({\em twisted}) cylinder, that have the same (inter)sections with a set of (two) cutting planes. 
 \noindent Hence, whatever is the reconstructed object from these sections, it would not be topologically consistent for at least one of these objects. In particular, the proposed reconstructed object ($\RR$) is a torus which is not homotopy equivalent to the ({\em twisted}) cylinder ($\OO$). In addition, we note that the Separation Condition may be verified for such a situation. Indeed, such a situation is exactly the case when the injective morphism between the fundamental groups of $\OO$ and $\RR$ is not surjective. This situation can be explained as follows: let $x_1$ and $x_2$ be two points in the sections $S_1$ and $S_2$ with the same lift $x$ in $\LSC$.  The lift of any curve $\gamma$ connecting $x_1$ and $x_2$ in $\MC$ provides a non-identity element of $\pi_1(\LSC,x)$ which is not in the image of $\LLL_*$. 
 We may avoid this situation with the following condition. 

\begin{figure}[!htb]
   \begin{center}  
\subfigure{\includegraphics[width=0.5\linewidth]{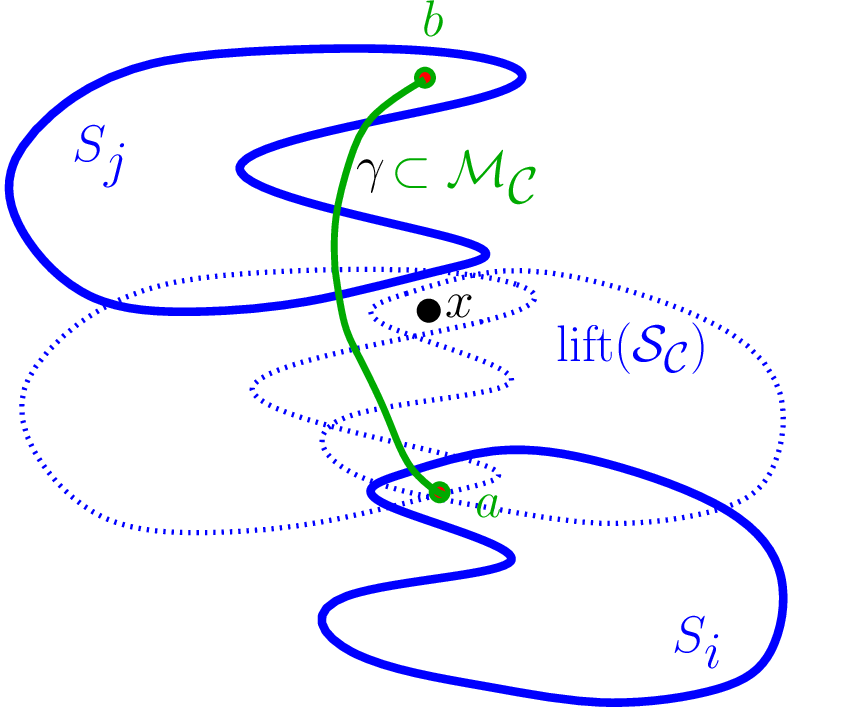}}
\caption{Intersection Condition.}
  \label{fig:int-condition}
  \end{center}
  \end{figure}

\begin{definition}[Intersection Condition] 
\rm We say that the set of cutting planes verifies the Intersection Condition if for any pair of sections $ S_i$ and $S_j$ in $\SC$, and for any connected component $X$ of $\lift(S_i) \cap \lift(S_j)$ (see Figure~\ref{fig:int-condition}), 
the following holds: there is a path $\gamma \subset \MC$ from a point $a\in S_i$ to a point $b \in S_j$ with $\lift(a) = \lift(b) =x \in X$ so that $\LLL_*(\gamma)$ is the identity element of  $\pi_1(\LSC,x)$, i.e., is contractible in $\LSC$ with a homotopy respecting the base point $x$. 
\end{definition}

In Section~\ref{sec:ensure}, we will show how to verify the Intersection Condition. Let us first prove the surjectivity of the map $\LLL_*$ which is deduced directly from the Intersection Condition.
\begin{theorem}[Surjectivity]
\label{th:part0}
 Under the Intersection Condition, the induced map $\LLL_*: \pi_1(\MC) \rightarrow \pi_1(\LSC)$ is surjective. 
\end{theorem}

\begin{proof}
Let $y_0$ be a fixed point of $\MC$ and $x_0 = \LLL(y_0)$. We show that $\LLL_* : \pi_1\bigl(\MC,y_0\bigr) \rightarrow \pi_1\bigl(\LSC,x_0\bigr)$ is surjective. Let $\alpha$ be a closed curve in $\LSC$ which represents an element of $\pi_1\bigl(\LSC,x_0\bigr)$. We show the existence of an element $\beta \in \pi_1\bigl(\MC,y_0\bigr)$ such that $\LLL_*(\beta) = [\alpha]$, where $[\alpha]$ denotes the homotopy class of $\alpha$ in $\pi_1\bigl(\LSC,x_0\bigr)$.
 \begin{figure}[!htb]
   \begin{center}
   \subfigure{\includegraphics[width=0.47\linewidth]{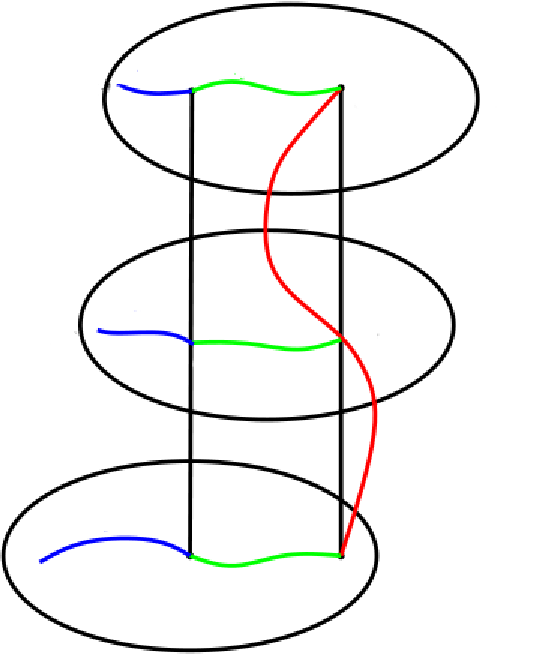}}
\caption{ For the proof of Theorem~\ref{th:part0}.}
  \label{fig:th6-new}
  \end{center}
  \end{figure}
\noindent We can divide $\alpha$ into subcurves $\alpha_1, \dots, \alpha_m$ such that $\alpha_j$ joins two points $x_{j-1}$ and $x_j$, and is entirely in the lift of one of the sections $S_{j}$, for $j=1, \dots, m$. We may assume $y_0 \in S_{1} = S_{m}$. For each $j=1,\dots, m$, let $\beta_j$ be the curve in $S_{j}$ joining two points $z_j$ to $w_j$ which is mapped to $\alpha_j$ under $\LLL$. Note that $w_j$ and $z_{j+1}$ (possibly) live in two different sections, but have the same image ($x_j$) under the lift map $\LLL$. Let $X_j$ be the connected component of $\lift(S_{j}) \cap \lift(S_{j+1})$ which contains $x_j$, see Figure~\ref{fig:th6-new}. 
According to the Intersection Condition, there is a path $\gamma_j \subset \MC$ connecting a point $a_j \in S_{j}$ to a point $b_{j+1} \in S_{{j+1}}$ such that $\lift(a_j) = \lift(b_{j+1}) = x'_j \in X_j$ and the image of $\gamma_j$ under $\LLL$ is the identity element of $\pi_1(\LSC,x'_j)$ (i.e., is contractible with a homotopy respecting the base point $x'_j$). Since $X_j$ is connected, there is a path from $x_j$ to $x'_j$ in $X_j$, so lifting back this path to two paths from $w_j$ to $a_j$ in $S_{j}$ and from $b_{j+1}$ to $z_{j+1}$ and taking the union of these two paths with $\gamma_j$, we infer the existence of a path $\gamma'_j \subset \MC$ connecting $w_j$ to $z_{j+1}$, such that the image of $\gamma'_j$ under $\LLL$ is contractible in $\LSC$ with a homotopy respecting the base point $x_j$.

Let $\beta$ be the path from $x_0$ to $x_0$ obtained by concatenating $\beta_j$ and $\gamma'_j$ alternatively, i.e., $\beta = \beta_1\gamma'_1\beta_2\gamma'_2\dots\beta_{m-1}\gamma'_m\beta_m\gamma'_m$. We claim that $\LLL_*([\beta]) = [\alpha]$. This is now easy to show: we have $\LLL_*(\beta) = \alpha_1\LLL_*(\gamma'_1)\alpha_2\dots\LLL_*(\gamma'_m)\alpha_m$, and all the paths $\LLL_*(\gamma'_j)$ are contractible to the constant path $[x_j]$ by a homotopy fixing $x_j$ all the time. We deduce that under a homotopy fixing $x_0$, $\alpha_1\LLL_*(\gamma'_1)\dots\LLL_*(\gamma'_m)\alpha_m \textrm{ is homotopic to }$  $\alpha_1\alpha_2\dots\alpha_m=\alpha,$  and this is exactly saying that $\LLL_*([\beta]) =[\alpha]$. The surjectivity follows. 
\end{proof}

\noindent Putting together all the materials we have obtained, we infer the main theorem of this section.
\begin{theorem}[Main Theorem-Part I]
Under the Separation and the Intersection Conditions, $\RC$ is homotopy equivalent to $\OC$ for any cell $\C$ of the arrangement.
\end{theorem}

\subsection{Generalized Nerve Theorem and homotopy equivalence of $\RR$ and $\OO$}
\label{subsec:nerve}

In this section, we extend the homotopy equivalence between $\RC$ and $\OC$, in each cell $\C$, to a global homotopy equivalence between $\RR$ and $\OO$. To this end, we make use of a generalization of the nerve theorem. This is a folklore theorem and has been observed and used by different authors. For a modern proof of a still more general result, we refer to Segal's paper~\cite{Segal}. (See also~\cite{May}, for a survey of similar results.)

\begin{theorem}[Generalized Nerve Theorem] 
 
Let $H:X \rightarrow Y$ be a continuous map. Suppose that $Y$ has an open cover $\mathcal K$ with the following two properties:
\begin{itemize}
\item Finite intersections of sets in $\mathcal K$ are in $\mathcal K$.
\item For each $U \in \mathcal K$, the restriction $H: H^{-1}(U) \rightarrow U$ is a weak homotopy equivalence.
\end{itemize}
Then $H$ is a weak homotopy equivalence.
\end{theorem}

Let $H_{\C}:\OC \rightarrow \RC$ be the homotopy equivalence obtained in the previous sections between $\OC$ and $\RC$. (So $H_{\C}$ is the composition of the retraction $\OC \rightarrow \MC$ and the inclusion $\MC \hookrightarrow \RC$.) Let $H: \OO \rightarrow \RR$ be the map defined by $H(x)= H_{\C}(x)$ if $x \in \OC$ for a cell $\C$ of the arrangement of the cutting planes. Note that $H$ is well-defined since $H_{\C}|_{\SC}= id_{\SC}$, for all $\C$. In addition, since for all cell $\C$, $H_{\C}$ is continuous, $H$ is continuous as well.

\noindent We can now apply the generalized nerve theorem by the following simple trick. Let $\epsilon$ be an infinitesimal positive value. For any cell $\C$ of the arrangement of the cutting planes, we define $\OC^{\epsilon}= \{x \in \R^3,\ d(x, \OC)< \epsilon \ \}$. Let us now consider the open covering $\mathcal K$ of $\OO$ by these open sets and all their finite intersections. It is straightforward to check that for $\epsilon$ small enough, the restriction of $H$ to each element of $\mathcal K$ is a weak homotopy equivalence. Therefore, according to the generalized nerve theorem, $H$ is a weak homotopy equivalence between $\RR$ and $\OO$.  And by Whitehead's theorem, $H$ is a homotopy equivalence between $\RR$ and $\OO$. Thus, we have

\begin{theorem}[Main Theorem-Part II]
{
Under the Separation and Intersection Conditions, the reconstructed object $\RR$ is homotopy equivalent to the unknown original shape $\OO$. 
}
\end{theorem}

\subsection{How to ensure the Intersection Condition?}
\label{sec:ensure}
We showed that by bounding from above the height of the cells by the reach of the object, we can ensure the Separation Condition.
In order to ensure the Intersection Condition, we need a stronger condition on the height of the cells. As we will see, this condition is a {\em transversality} condition on the cutting planes that can be measured by the angle between the cutting planes and the normal to $\partial \OO$ at contour-points. 

\begin{figure}[!htb]
 \begin{center}
\subfigure{\includegraphics[width=0.4\linewidth]{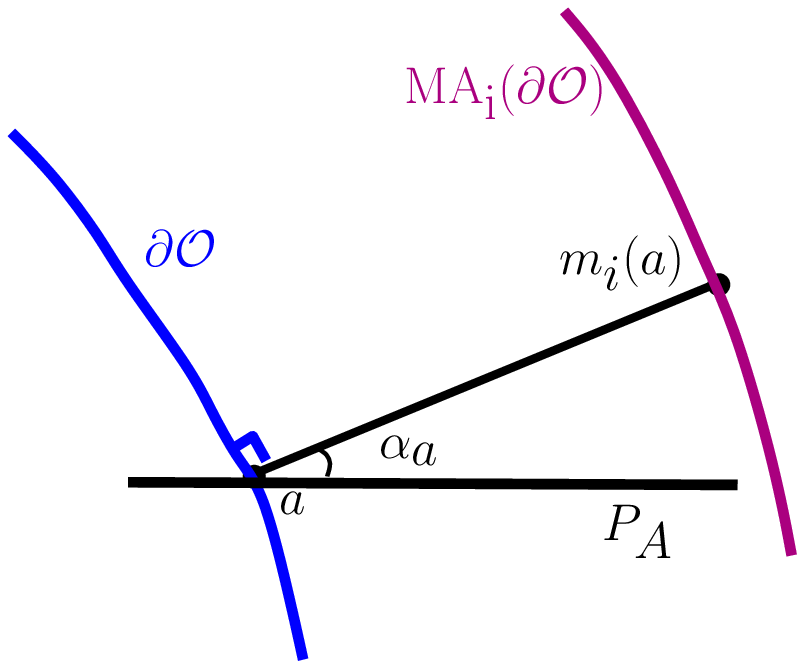}}
 \caption{Definition of $\alpha_a$ (a 2D illustration).}
 \label{fig:alpha}
 \end{center}
 \end{figure}

\begin{definition}[Angle $\alpha_a$]
\rm Let $a$ be a point on the boundary of a section $A \in \SC$ on the plane $P_A$. We consider $m_i(a)$, that may be outside the cell $\C$. We define $\alpha_a$ as the angle between $P_A$ and the normal to $\partial \OO$ at $a$, i.e., $\alpha_a := \mathrm{angle}(P_A,[a,m_i(a)]),$ see Figure~\ref{fig:alpha}.
We define $\alpha_{\C}=\max_{a \in \partial \SC} \alpha_a.$

\end{definition}

\paragraph*{\bf Sufficient conditions}  We now define the sampling conditions on the cutting planes. (See Section \ref{sec:ensure-separation} for the definitions of $h_{\C}$ and $\reach_{\C}(\OO)$.)

\begin{description}
\item[(C1) Density Condition] 

For any cell $\C$ of the arrangement, $h_{\C} < \reach_{\C}(\OO).$

\item[(C2) Transversality Condition]

For any cell $\C$, $$h_{\C} < \frac{1}{2} \; \bigl(1 - \sin(\alpha_{\C})\bigr) \; \reach_{\C}(\OO). $$
\end{description}

\noindent The Density Condition is based on the density of the sections. 
The Transversality Condition is defined in a way that the transversality of the cutting planes to $\partial \OO$ and the distance between the sections are controlled simultaneously. (Indeed, $\sin(\alpha_{\C})$ is to control the transversality, and bounding from above $h_{\C}$ allows us to control the distance between the sections.) Note that the Transversality Condition implies trivially the Density Condition.

\subsubsection*{Remarks} The transversality of the cutting planes to $\partial \OO$ seems to be a reasonable condition in practice, specially for applications in 3D ultrasound. Indeed, according to \cite{R03} Section 1.2.1, from a technical point of view if a cut is not sufficiently transversal to the organ, the quality of the resulting 2D ultrasonic image is not acceptable for diagnosis.
\vspace{2mm}

The following theorem is the main result of this paper:


\begin{theorem}[Main Theorem-Part III]
\label{th:PartIII}
{
If the set of cutting planes verifies the Transversality Condition (and so the Density Condition), then the Separation and the Intersection Conditions are verified. Therefore, the proposed reconstructed object $\RR$ is homotopy equivalent to the unknown original shape $\OO$. 
}
\end{theorem}
\subsection{\em Proof of Theorem~\ref{th:PartIII}}
\begin{figure}[!htb]
 \begin{center}
\subfigure{\includegraphics[width=0.5\linewidth]{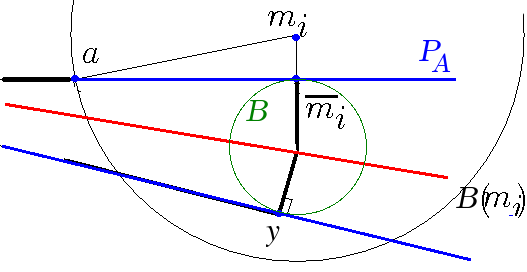}}
 \caption{Definition of $\bar{m}_i(a)$ (a 2D illustration), and for the proof of Lemma~\ref{K_i}.}
 \label{fig:bar-mi}
 \end{center}
 \end{figure}

\noindent This section is devoted to the proof of Theorem~\ref{th:PartIII}. We need the following notations. 
\paragraph*{\bf Notation $\bigl( K_i(\SC)$ and $K_e(\SC)\bigr)$} Recall that $\VD(\C)$ is the locus of the points with more than one nearest point in $\partial \C$. We write $K_i(\SC)$ (resp. $K_e(\SC)$) for the set of points $x \in \VD(\C)$ such that all the nearest points of $x$ in $\partial \C$ lie inside (resp. outside) the sections. 
\paragraph*{\bf Notation $\bigl( \bar{m}_i(a)$ and $\bar{m}_e(a)\bigr)$} Let $a$ be a point on the boundary of a section $A \in \SC$ on the plane $P_A$.
We write $\bar{m}_i(a)$ (resp. $\bar{m}_e(a)$) for the orthogonal projection of $m_i(a)$ (resp. $m_e(a)$) onto $P_A$. 
See Figure~\ref{fig:bar-mi}. We have $d(m_i(a), \bar{m}_i(a))= \sin(\alpha_a) \; d(a, m_i(a))$ and $d(m_e(a), \bar{m}_e(a))= \sin(\alpha_a) \; d(a, m_e(a))$.

 \begin{lemma}
\label{K_i} \rm
If the Transversality Condition is verified, for any $a \in \partial \SC$, we have 
$\lift(\bar{m}_i(a)) \in K_i(\SC)$ and $\lift(\bar{m}_e(a)) \in K_e(\SC).$ In addition, the two segments that join $\lift(\bar{m}_i(a))$ to its nearest points in $\partial \C$ both lie entirely in $\MC$.
 \end{lemma}
\begin{proof}
{\rm
We first show that $\lift(\bar{m}_i(a)) \in K_i(\SC)$. The symmetric property for $\lift(\bar{m}_e(a))$ can be proved similarly. Let us simplify the notation by writing $m_i$ for $m_i(a)$, and $\bar{m}_i$ for $\bar{m}_i(a)$. 
 Let us also write $B(m_i)$ for the ball centered at $m_i$ of radius $d(m_i, a)$. We have $d(m_i, \bar{m}_i) \leq d(m_i, a)$. Thus, $\bar{m}_i$ lies in $B(m_i)$. As this ball is contained in $\OO$, $\bar{m}_i$ is in $\OO$. Consider now $\lift(\bar{m}_i)$ on $\VD(\C)$, and call $y$ the point distinct from $\bar{m}_i$ such that $\lift(\bar{m}_i)=\lift(y)$, see Figure~\ref{fig:bar-mi}. 
 To have $\lift(\bar{m}_i) \in K_i(\SC)$, we show that $y$ is in $\OO$. We have: $$d(m_i, y) \leq d(m_i, \bar{m}_i) + d(\bar{m}_i, \lift(\bar{m}_i)) + d(\lift(y), y)$$ $$\leq \sin(\alpha_a) d(a, m_i) + 2 \; h_{\C} \leq d(a, m_i).$$ Thus, $y$ belongs to $B(m_i)$, and we can deduce that $y \in \OO$ and $\lift(\bar{m}_i) \in K_i(\SC)$. Consider now the ball $B$ centered at $\lift(\bar{m}_i)$ which passes through $\bar{m}_i$ and $y$. We claim that $B$ is entirely contained in $B(m_i)$. Since for any point $p \in B$ we have:
$$d(m_i, p) \leq d(m_i, \lift(\bar{m}_i)) + d(\lift(\bar{m}_i), p)\leq \sin(\alpha_a) d(a, m_i) + 2 \; h_{\C} \leq d(a, m_i).$$

\noindent Thus, $B \subset B(m_i) \subset \OO$ and the interior of $B$ is empty of points of $\partial \OC$. Therefore, $B$ is a medial ball of $\OC$, and its center $\lift(\bar{m}_i)$ belongs to $\MA_i(\partial \OC)$. Since $\bar{m}_i$ and $y$ are in $\SC$, according to the definition of $\MC$, the line-segments $[\lift(\bar{m}_i),\bar{m}_i]$ and $[\lift(\bar{m}_i),y]$ lie entirely in $\MC$.
}
\end{proof}

\begin{lemma}
\label{lemma:Ensure-Int-con}
{\rm
Under the Transversality Condition, the Intersection Condition is verified.
}
\end{lemma}

\begin{proof}
{\rm 
Let $S_a$ and $S_b$ be two sections in $\SC$ such that $\lift(S_a) \cap \lift(S_b)$ is non-empty. We recall that $\LLL_*$ denotes the homomorphism $\LLL_*: \pi_j(\MC) \rightarrow \pi_j(\LSC)$ induced by the lift function $\LLL: \MC \rightarrow \LSC$. 
 We will show that for any two points $a \in \partial S_a, b \in \partial S_b$ such that $\lift(a)=\lift(b)$, there exists a path $\gamma \subset \MC$ between $a$ and $b$ so that $\LLL_*(\gamma)$ is contractible in $\LLL_*(\pi_1(\MC))$. Let $P_a$ and $P_b$ be the cutting-planes of $S_a$ and $S_b$ respectively. One of the two following cases happen:

\medskip

\noindent {\bf Case 1:}  If the segment $[a,\bar{m}_i(a)]$ (i.e., the projection of the segment $[a,m_i(a)]$ onto $P_a$) is not cut by any other cutting plane, then we claim that $\lift(a)$ is connected to $\lift(\bar{m}_i(a))$ in $K_i(\SC)$.

\noindent Consider the 2-dimensional disk $B$ in the plane $P_a$ centered at $\bar{m}_i(a)$ and passing through $a$. This disk is contained in the 3D ball centered at $m_i(a)$ and passing through $a$, which lies entirely inside $\OO$. Considering the intersection of this 3D ball with $P_a$, we infer that $B$ lies inside section $S_a$. Therefore, $\lift(B)$ is entirely contained in $\lift(S_a)$. In Figure~\ref{fig:l4-bis}, $\lift(B)$ is colored in green.
 \begin{figure}[!htb]
   \begin{center}
     \subfigure{\includegraphics[width=0.5\linewidth]{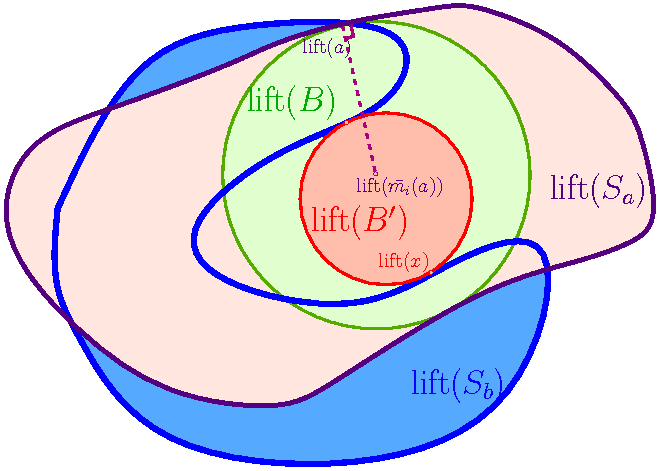}}
\caption{ For the proof of Lemma~\ref{lemma:Ensure-Int-con}: the lift of two sections $S_a$ and $S_b$ on $\VD(\C)$ is illustrated. We prove that the lift of the segment $[a,\bar{m}_i(a)]$ lies in $\lift(S_a) \cap \lift(S_b)$.} 
  \label{fig:l4-bis}
  \end{center}
  \end{figure} 

\noindent Since $\lift(a)=\lift(b)$, we have $\lift(a) \in \lift(S_a) \cap \lift(S_b)$. On the other hand, according to Lemma~\ref{K_i}, $\lift(\bar{m}_i(a))$ lies in $\lift(S_a) \cap \lift(S_b)$. For the sake of contradiction, suppose that $\lift(a)$ is not connected to $\lift(\bar{m}_i(a))$ in $K_i(\SC)$.
In this case, $\lift(B)$ intersects (at least) two different connected components of $\lift(S_a) \cap \lift(S_b)$, see Figure~\ref{fig:l4-bis}. Since $\lift(B) \subset \lift(S_a)$, we deduce that $\lift(B)$ intersects $\lift(S_b)$ in (at least) two different connected components. 
Then, consider the maximal open disk contained in $\lift(B)$ which is empty of points of $\lift(S_b)$. Such a disk consists of the lift of a 2D disk $B'$ in the plane $P_b$ which is tangent to $\partial S_b$ at two points $x$ and $x'$. $B'$ is the intersection of the 3D medial ball of the complementary set of $\OO$ passing through $x$ and $x'$. Thus  $\bar{m}_e(x)$, which is the projection of the center of this 3D ball ($m_e(x)$) onto $P_b$, lies in $B'$.
We have $\lift(\bar{m}_e(x)) \in \lift(B') \subseteq \lift(B) \subset \lift(S_a)$. Thus, one of the nearest points of $\lift(\bar{m}_e(x))$ in $\partial \C$ lies in $S_a \subset \SC$. This contradicts $\lift(\bar{m}_e(a)) \in K_e(\SC)$ (Lemma~\ref{K_i}).

\noindent Therefore, $\lift(a)$ is connected to $\lift(\bar{m}_i(a))$ in $K_i(\SC)$. Let us call $a'$ and $b'$ the nearest points of $\lift(\bar{m}_i(a))$ in $S_a$ and $S_b$ respectively, see Figure~\ref{fig:l4}-left. According to Lemma~\ref{K_i}, the line-segments $[a',\lift(\bar{m}_i(a))]$ and $[b',\lift(\bar{m}_i(a))]$ lie inside $\MC$. We now define a path $\gamma$, colored in red in Figure~\ref{fig:l4}-left, as the concatenation of four line-segments : $[a,a'] \subset S_a$, $[a',\lift(\bar{m}_i(a))]$, $[b',\lift(\bar{m}_i(a))]$ and $[b',b] \subset S_b$. We know that $[a',\lift(\bar{m}_i(a))]$ and $[b',\lift(\bar{m}_i(a))]$ are mapped to a single point $\lift(\bar{m}_i(a))$ by the lift function. Thus, the image of $\gamma$ under the lift function is the line-segment $[\lift(a), \lift(\bar{m}_i(a))]$, which is trivially contractible in $\LSC$. 

 \begin{figure}[!htb]
   \begin{center}
     \subfigure{\includegraphics[width=0.31\linewidth]{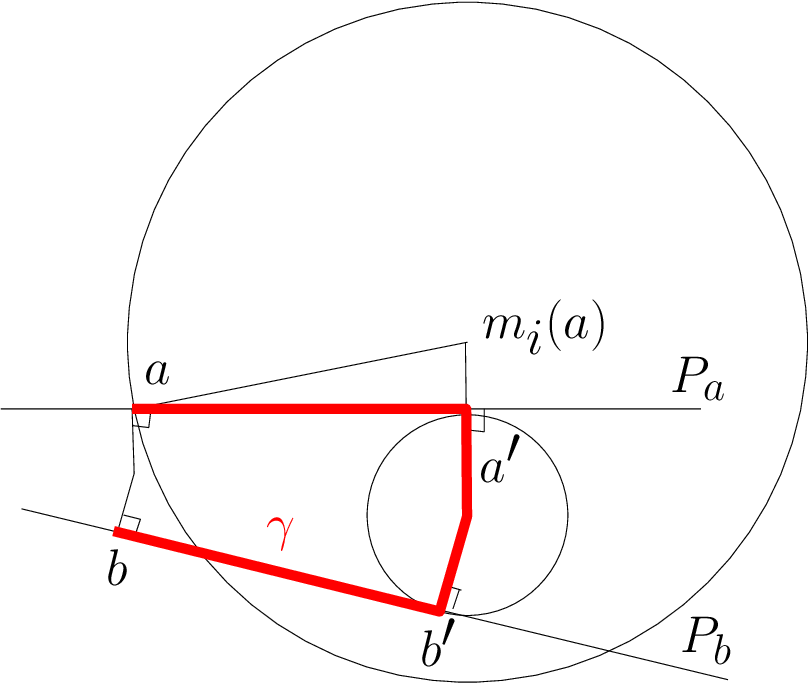}}
\hspace{2mm} 
    \subfigure{\includegraphics[width=0.31\linewidth]{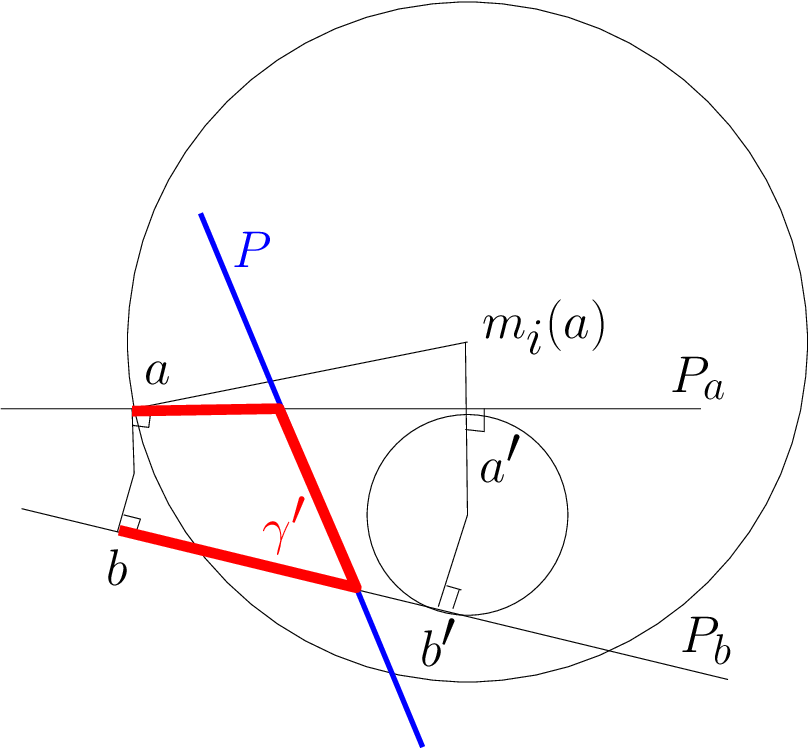}}
\hspace{2mm} 
    \subfigure{\includegraphics[width=0.31\linewidth]{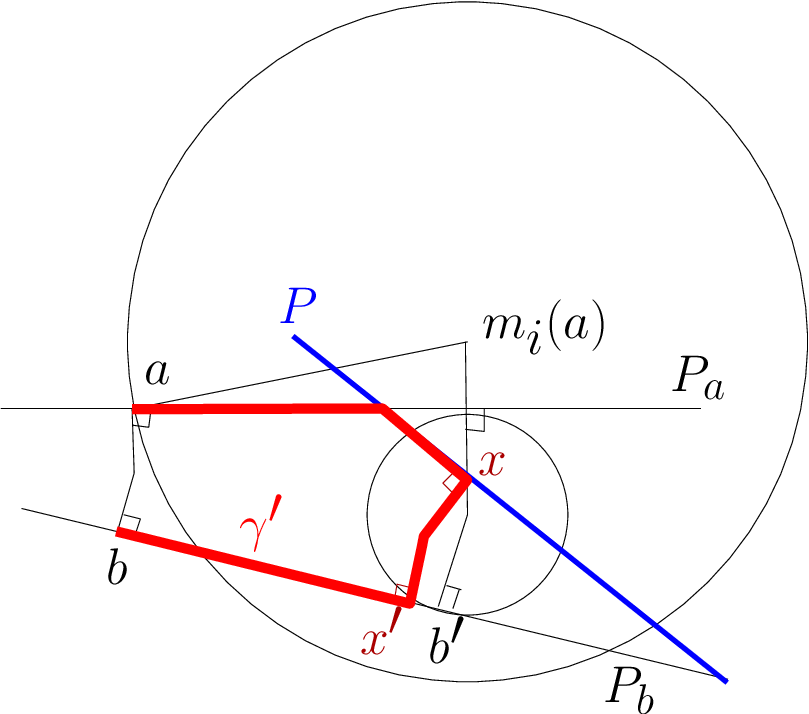}}
\caption{ A 2D illustration of the cases considered in the proof of Lemma~\ref{lemma:Ensure-Int-con} : in red a path between $a$ and $b$ in $\MC$.} 
  \label{fig:l4}
  \end{center}
  \end{figure}

 \noindent {\bf Case 2:} We now consider the case where the segment $[a,\bar{m}_i(a)]$ is cut by a cutting plane $P$ (colored in blue in Figure~\ref{fig:l4}-~middle and -right). If $\C$ is the cell which contains $a$ and $b$, the path $\gamma=[aa'b'b]$ (defined above) does not entirely lie in $\C$. Let us consider the intersection of $\partial \C$ with the plane of $\gamma$ denoted by $P_0$ (the plane of Figure~\ref{fig:l4}). Two cases may happen:
\begin{itemize}
\item If $\partial \C$ does not intersect $[a',\lift(\bar{m}_i(a))]$ and $[b',\lift(\bar{m}_i(a))]$, we define $\gamma':=\partial \C \cap P_0$ which is a path that connects $a$ to $b$. See Figure~\ref{fig:l4}-middle. Note that $\gamma'$ is a path in $\MC$, since it lies in $\partial \C \cap \OO= \SC$. 
\item Otherwise, let $x$ be an intersection point (the closest to $a$) between $\partial \C$ and $[a',\lift(\bar{m}_i(a))]\cup [b',\lift(\bar{m}_i(a))]$, see Figure~\ref{fig:l4}-right. Let $x'$ be so that $\lift(x)=\lift(x')$. We now define a path $\gamma' \subset \MC$ which connects $a$ to $b$ along $\partial \C \cap P_0$, while taking a shortcut from $x$ to $x'$ (by $[x,\lift(x)]$ and $[\lift(x),x']$). This path is colored in red in Figure~\ref{fig:l4}-right. Note that in the case $\partial \C \cap P_0$ is formed of more planes, $\gamma'$ can be defined in a similar way by following $\partial \C \cap P_0$ and making a shortcut at each intersection with $\gamma$, i.e. $x$ in our example. 
\end{itemize}
We now claim that (for both cases considered above) the lift of $\gamma'$ is contractible in $\LSC$. Indeed, the lift of the first and the last line-segments of $\gamma'$, that lie respectively on $P_a$ and $P_b$, is a line segment. Therefore, it is sufficient to check the contractibility of the lift of the part of $\gamma'$ which is between $P_a$ and $P_b$. The line-segments of $\gamma'$ that connect $P_a$ to $P_b$, lie on sections that are all entirely contained in the medial ball centered at $m_i(a)$. Therefore, the lift of these segments is contained in a disk and is contractible in $\LSC$.

}
\end{proof}
This ends the proof of Theorem~\ref{th:PartIII}.
\subsection{Deforming the homotopy equivalence to a homeomorphism}\label{sec:homeomorphism}
Using the homotopy equivalence between $\RR$ and $\OO$, we now show that they are indeed isotopic.
\begin{theorem}[Main Theorem-Part IV]
Under the Separation and the Intersection Conditions, the two topological manifolds $\RR$ and $\OO$ are homeomorphic (in addition, they are isotopic).
\end{theorem}
Although this result is stronger than the homotopy equivalence, the way our proof works makes essentially use of the topological study of the previous sections. 

\paragraph*{\bf Proof} Again, we first argue in each cell of the arrangement and show the existence of a homeomorphism between $\OC$ and $\RC$ whose restriction to $\SC$ is the identity map. 
Gluing these homeomorphisms together, one obtains a global homeomorphism between $\RR$ and $\OO$.  Let $\C$ be a cell of the arrangement of the cutting planes. A similar method used to prove the homotopy equivalence between $\RC$ and $\OC$ shows that $\partial \RR \cap \C$ and $\partial \OO \cap \C$ are homotopy equivalent 2-manifolds and are therefore homeomorphic, and in addition there exists a homeomorphism $\beta_\C :  \partial \OO \cap \C \rightarrow \partial \RR \cap \C$ which induces identity on the boundary of sections in $\SC$. We showed that the topology of  $\RC$ and $\OC$ is completely determined by their fundamental groups, i.e., all the higher homotopy groups of $\RC$ and $\OC$ are trivial. Moreover, there is an isomorphism between $\pi_1(\OC)$ and $\pi_1(\RC)$, and the induced map $(\beta_\C)_*:\pi_1(\partial \OO \cap \C) \rightarrow \pi_1(\partial \RR \cap \C)$ on the first homotopy groups is consistent with this isomorphism (in the sense that there exists a commutative diagram of first homotopy groups).  This shows that $\beta_\C$ can be extended  to a map $\alpha_\C : \OC \rightarrow \RC$, inducing the corresponding isomorphism between $\pi_1(\OC)$ and $\pi_1(\RC)$, and such that the restriction of $\alpha_\C$ to $\SC$ remains identity. Since all the higher homotopy groups of $\OC$ and $\RC$ are trivial, it follows that $\alpha_\C$ is a homotopy equivalence. 
We can now apply the following theorem due to Waldhausen which shows that $\alpha$ can be deformed to homeomorphism between $\OC$ and $\RC$ by a deformation which does not change the homeomorphism $\alpha_\C$ between the boundaries. A compact 3-manifold $M$ is called {\it irreducible} if $\pi_2(M)$ is trivial. We note that $\OC$ and $\RC$ are irreducible.
\begin{theorem}[Waldhausen]

Let $f:M \rightarrow M'$ be a homotopy equivalence between orientable irreducible
3-manifolds with boundaries such that $f$ takes the boundary of $M$ onto the
boundary of $M'$ homeomorphically. Then $f$ can be deformed to a homeomorphism
$M \rightarrow M'$ by a homotopy which is fixed all the time on the boundary of $M$.
 \end{theorem}
\noindent (See \cite{Matveev03}, page 220, for a proof.)
\medskip

Applying Waldhausen's theorem, one obtains a homeomorphism $\tilde\alpha_\C$ from $\OC$ to $\RC$ which is identity on the sections in $\SC$. Gluing $\tilde\alpha_C$, one  obtains a global homeomorphism from $\OO$ to $\RR$.
Moreover, according to Chazal and Cohen-Steiner's work~\cite{CC05} (Corollary 3.1), since $\RR$ and $\OO$ are homeomorphic and $\RR$ contains the medial axis of $\OO$, $\RR$ is isotopic to $\OO$.  \qquad \hspace*{\fill}$\Box$

\subsection*{Conclusion}
In this paper, we presented the first topological studies in shape
reconstruction from cross-sectional data.  We showed that the generalization of the
classical overlapping criterion to solve the correspondence problem between unorganized cross-sections
preserves the homotopy type of the shape under some appropriate sampling conditions. In addition, we proved that in this case, the homotopy equivalence between the reconstructed object and the original shape can be deformed to a homeomorphism. Even, more strongly, the two objects are isotopic. 

As a future work, higher dimensional variants of this reconstruction problem can be considered where the goal will be to reconstruct an $n$-dimensional shape from its $(n-1)$-dimensional intersections with hyperplanes. 
Similar analysis can be carried on for the resulting $n$-dimensional reconstruction: The Separation Condition can be defined in any dimension. Under this condition, the guarantees on the connectivity between the sections remain valid. 
Using the lift function, we can similarly define a map between $\OC$ and $\RC$. In addition, we can see that under the Separation Condition, this map induces injective morphisms between the homotopy groups of $\OC$ and $\RC$. The main issue then will be to impose the surjectivity at the level of higher homotopy groups of $\OC$ and $\RC$.
Once the surjectivity is verified under some new appropriate sampling conditions on the cutting planes, the homotopy equivalence between $\OO$ and $\RR$ will follow.

\subsection*{Acknowledgments}
This work has been partially supported by the High Council for Scientific and Technological Cooperation between Israel and France (research networks program in medical and biological imaging).  The authors would like to thank David Cohen-Steiner for helpful discussions, and Dominique Attali, Gill Barequet and Andr\'e Lieutier for their careful reading of the manuscript and their constructive comments.

\bibliographystyle{alpha}
\bibliography{reconstruction}

\appendix
\section{Homotopy Theory Basics}
\label{sec:Homotopy-Preliminaries}
\noindent In this section, we briefly recall some basic definitions an results that are used in this paper.
 
\begin{definition} [Homotopy]
{\rm
A {\em homotopy} between two continuous functions $f$ and $g$ from a topological space $X$ to a topological space $Y$ is defined to be a continuous function $H: X \times [0,1] \rightarrow Y$ such that for all points $x \in X$, $H(x,0)=f(x)$ and $H(x,1)=g(x)$. $f$ is said to be {\em homotopic} to $g$ if there exists a homotopy between $f$ and $g$.
}
\end{definition}

\begin{definition} [Homotopy Equivalence]
{\rm Two topological spaces $X$ and $Y$ are of the {\em same homotopy type} ({\em homotopy equivalent}) if there exist continuous maps $f: X \rightarrow Y$ and $g: Y \rightarrow X$ such that $g \circ f$ is homotopic to the identity map $id_X$ and $f \circ g$ is homotopic to $id_Y$.

%
}
\end{definition}

\begin{definition}[Homotopy Groups, Fundamental Group]
\rm Let $X$ be a space with a base point $x_0 \in X$. Let $S^i$ denote the $i$-sphere for a given $i \geq 1$, in which we fixed a base point $b$. The $i$-dimensional homotopy group of $X$ at the base point $x_0$, denoted by $\pi_i(X,x_0)$, is defined to be the set of homotopy classes of maps $f: S^i \rightarrow X$ that map the base point $b$ to the base point $x_0$.

\noindent If $X$ is path-connected, the group $\pi_i(X,x_0)$ is, up to isomorphism, independent
of the choice of base point $x_0$. In this case the notation $\pi_i(X,x_0)$ is often
abbreviated to $\pi_i(X)$. 

Let $X$ be a path-connected space. The first homotopy group of $X$, $\pi_1(X)$, is called the {\em fundamental group} of $X$. 
\end{definition}
\begin{definition}[Simply-Connected Spaces]
\rm
The path-connected space $X$ is called {\em simply-connected} if its fundamental group is trivial.
\end{definition}

\begin{definition}[Weak Homotopy Equivalence]
\rm A map $f:X \rightarrow Y$ is called a weak homotopy equivalence if the group homomorphisms induced by $f$ on the corresponding homotopy groups, $f_*: \pi_i(X) \rightarrow \pi_i(Y)$, for $i\geq 0$, are all isomorphisms.
It is easy to see that any homotopy equivalence is a weak homotopy equivalence, but the inverse is not necessarily true in general.
However, Whitehead's Theorem states that the inverse is true for maps between CW-complexes.
\end{definition}
\begin{theorem}[Whitehead's Theorem]
\rm If a map $f:X \rightarrow Y$ between connected CW-complexes induces isomorphisms $f_*: \pi_i(X) \rightarrow \pi_i(Y)$ for all $i\geq 0$, then $f$ is a homotopy equivalence.
\end{theorem}

\begin{definition}[(Strong) Deformation Retract]
\rm Let $X$ be a subspace of $Y$. A homotopy $H: Y \times [0,1] \rightarrow Y$ is said to be a (strong) deformation retract of $Y$ onto $X$ if:
\begin{itemize}
\item For all $y \in Y, \ H(y,0)=y$ and $H(y,1) \in X$.
\item For all $x \in X, H(x,1)=x$.
\item (and for all $x \in X, H(x,t)=x$.)
\end{itemize}
\end{definition}

\begin{definition} [Homeomorphism]
{\rm
Two topological spaces $X$ and $Y$ are {\em homeomorphic} if there exists a continuous and bijective
map $h: X \rightarrow Y$ such that $h^{-1}$ is continuous. the map $h$ is called a homeomorphism from $X$ to $Y$.
}\end{definition}

\begin{definition} [Isotopy]
{\rm
Two topological spaces $X$ and $Y$ embedded in $\R^d$ are {\em isotopic} if there exists a continuous map $i: [0,1] \times X \rightarrow \R^d$ such that $i(0,.)$ is the identity over $X$, $i(1,X)=Y$ and for any $t \in [0,1]$, $i(t,.)$ is a homeomorphism from $X$ to its image. The map $i$ is called an isotopy from $X$ to $Y$.
}\end{definition}

\noindent Let us recall the definition of the universal cover, and refer to classical books in topology for more details.

\begin{definition}[Universal Cover]
\rm Let $X$ be a topological space. A {\em covering space} of $X$ is a space $C$ together with a continuous surjective map $\phi: C \rightarrow X$ such that for every $x \in X$, there exists an open neighborhood $U$ of $x$, such that $\phi^{-1}(U)$ is a disjoint union of open sets in $C$, each of which is mapped homeomorphically onto $U$ by $\phi$. A connected covering space is called a {\em universal cover} if it is simply connected. The universal cover exists and is unique up to isomorphism of covering spaces.
\end{definition}
\begin{lemma}[Lifting Property of the Universal Cover]
\rm Let $X$ be a (path-) connected topological space and $\widetilde{X}$ be its universal cover, and $\phi: \widetilde{X} \rightarrow X$ be the map given by the covering. Let $Y$ be any simply connected space, and $f: Y \rightarrow X$ be a continuous map. Given two points $\tilde x\in\tilde X$ and $y \in Y$ with $\phi(\tilde x)=f(y)$, there exists a unique continuous map $g: Y \rightarrow \widetilde{X}$ so that $\phi \circ g=\ f$ and $\phi(y)=\tilde x$. 
 This is called the {\em lifting property} of $\widetilde{X}$. 
\end{lemma}
Since all the spheres $S_i$ of dimensions $i\geq 2$ are simply connected, we may easily deduce
\begin{corollary}\label{cor:univ}
\rm For any (path-) connected space $X$ with universal cover $\tilde X$, we have $\pi_i(\tilde X)=\pi_i(X)$ for all $i\geq 2$.
\end{corollary}

\noindent We now recall the definition of the Hurewicz map $h_i : \pi_i(X) \rightarrow H_i(X)$. For an element $[\alpha] \in \pi_i(X)$ presented by $\alpha: S^i \rightarrow X$, $h_i([\alpha])$ is defined as the image of the fundamental class of $S^i$ in $H_i(S^i)$ under the map $\alpha_*: H_i(S^i) \rightarrow H_i(X)$, i.e., $h_i([\alpha])= \alpha_*(1)$. 
\begin{theorem}[Hurewicz Isomorphism Theorem]
\rm
The first non-trivial homotopy and homology groups of a simply-connected space occur in the same dimension and are isomorphic. In other words,  for $X$ simply connected, the Hurewicz map $h_i : \pi_i(X) \rightarrow H_i(X)$ is an isomorphism for the first $i$ with $\pi_i$ (or equivalently $H_i$) non-trivial.
\end{theorem}

\end{document}